\documentclass[11pt]{article}
\usepackage{natbib}
\usepackage{amssymb,amsmath,amsthm,latexsym,amsfonts,amscd,dsfont,enumerate}
\usepackage{a4wide,graphicx,tikz}

\usetikzlibrary{arrows,shapes,shadows,fit,backgrounds}
\tikzstyle{trader} = [circle, draw, top color=white, bottom color=blue!30, draw=blue!50!black!100, drop shadow, minimum height=4em]
\tikzstyle{bank} = [rectangle, draw, top color=white, bottom color=red!20, draw=red!50!black!100, drop shadow, rounded corners, minimum height=3em, text width=4em, text centered]
\tikzstyle{market} = [rectangle, draw, top color=white, bottom color=green!20, draw=green!50!black!100, drop shadow, rounded corners, minimum height=3em, text width=4em, text centered]
\tikzstyle{background} = [rectangle,fill=gray!10, inner sep=0.2cm, rounded corners=5mm]
\tikzstyle{line} = [draw, latex'-latex']
\tikzstyle{from} = [draw, latex'-]
\tikzstyle{to} = [draw, -latex']

\usepackage[]{hyperref}
 \hypersetup{
 colorlinks=true, 
 breaklinks=true, 
 urlcolor= blue, 
 linkcolor= blue, 
 bookmarksopen=true, 
 pdfauthor={Pallavicini, A., Brigo D.}
 }

\newtheorem{theorem}{Theorem}[section]
\newtheorem{remark}[theorem]{Remark}

\newcommand{\ExG}[2]{\mathbb{E}\!\left[\,#2\,\left|{\cal G}_{#1}\right.\right]}

\newcommand{\ExFT}[3]{\mathbb{E}^{#2}\!\left[\,#3\,\left.\right|{\cal F}_{#1}\right]}
\newcommand{\ExGT}[3]{\mathbb{E}^{#2}\!\left[\,#3\,\left.\right|{\cal G}_{#1}\right]}

\newcommand{\ExT}[3]{\mathbb{E}_{#1}^{#2}\!\left[\,#3\,\right]}

\newcommand{\PxC}[2]{\mathbb{P}\!\left\{\,#1\,\left.\right|\,#2\,\right\}}

\newcommand{\QxCT}[3]{\mathbb{Q}^{#1}\!\left\{\,#2\,\left.\right|\,#3\,\right\}}

\newcommand{\ind}[1]{1_{\{#1\}}}

\newcommand{\VxFT}[3]{{\rm Var}^{#2}\!\left[\,#3\,\left.\right|{\cal F}_{#1}\right]}

\newcommand{\rec}{R}
\newcommand{\lgd}{\mbox{L{\tiny GD}}}

\newcommand{\cva}{\mbox{CVA}}
\newcommand{\dva}{\mbox{DVA}}
\newcommand{\fva}{\mbox{FVA}}
\newcommand{\mva}{\mbox{MVA}}
\newcommand{\mtm}{\mbox{MtM}}

\newcommand{\beq}[0]{\begin{equation}}
\newcommand{\eeq}[0]{\end{equation}}
\newcommand{\beqn}[0]{\begin{equation*}}
\newcommand{\eeqn}[0]{\end{equation*}}
\newcommand{\balign}[0]{\begin{aligned}}
\newcommand{\ealign}[0]{\end{aligned}}

\title{CCP Cleared or Bilateral CSA Trades with Initial/Variation Margins under  credit, funding and wrong-way risks:\\ A Unified Valuation Approach\thanks{The opinions here expressed are solely those of the authors and do not represent in any way those of their employers. This paper is a refined version of the initial report \cite{Pallavicini2013z}}
}
\author{
Damiano Brigo\thanks{Imperial College London and Capco Institute, {\tt damiano.brigo@imperial.ac.uk}}
\ \ \
Andrea Pallavicini\thanks{Imperial College London and Banca IMI Milan, {\tt a.pallavicini@imperial.ac.uk}}
}
\date{
\small First Version: July 9, 2013.  This version: \today
}

\begin{document}

\maketitle

\begin{abstract}
The introduction of CCPs in most derivative transactions will dramatically change the landscape of derivatives pricing, hedging and risk management, and, according to the TABB group, will lead to an overall liquidity impact about 2 USD trillions. In this article we develop for the first time a comprehensive approach for pricing under CCP clearing, including variation and initial margins, gap credit risk and collateralization, showing concrete examples for interest rate swaps. Mathematically, the inclusion of asymmetric borrowing and lending rates in the hedge of a claim lead to nonlinearities showing up in claim dependent pricing measures, aggregation dependent prices, nonlinear PDEs and BSDEs. This still holds in presence of CCPs and CSA. We introduce a modeling approach that allows us to enforce rigorous separation of the interconnected nonlinear risks into different valuation adjustments where the key pricing nonlinearities are confined to a funding costs component that is analyzed through numerical schemes for BSDEs. We present a numerical case study for Interest Rate Swaps that highlights the relative size of the different valuation adjustments and the quantitative role of initial and variation margins, of liquidity bases, of credit risk, of the margin period of risk and of wrong way risk correlations. 
\end{abstract}

\medskip

\noindent\textbf{AMS Classification Codes}: 62H20, 91B70 \newline

\medskip

\noindent \textbf{Keywords}: Central Counterparty Clearing, Credit Support Annex, ISDA, Interest Rate Derivatives, Funding Costs, Bilateral Counterparty Risk, Credit Valuation Adjustment, CVA, Collateral Modeling, Initial Margin, Variation Margin, Close-Out, Re-hypothecation, Gap Risk, Margin Period of Risk, Backward Stochastic Differential Equations, Funding Valuation Adjustment, Margin Valuation Adjustment.

\newpage
{\small \tableofcontents}
\newpage

\pagestyle{myheadings} \markboth{}{{\footnotesize D. Brigo and A. Pallavicini: A Unified Pricing Framework}}

\section{Introduction}
\label{sec:introduction}


In this paper we face the problem of designing a quantitative methodology that may be applied to valuation of trades occuring under central counterparties (CCP) clearing or via a bilateral Credit Support Annex (CSA). A crucial element of this analysis is the inclusion of the relevant risks affecting the trades besides their standard market risks, namely credit risk, wrong way risk and collateral gap risk, initial and variation margins, funding costs and default modeling. While this paper focuses on the analysis at deal level of a specific market player, namely what we could call the micro view, the onset of CCPs is a relevant issue also from the systemic / macro point of view. In this respect, it is appropriate to mention that in the large scale space the Tabb Group estimated extra collateral requirements following the full onset of CCPs to hit a level of 2 $\$$ Trillion, see \cite{Tabb2011}. This figure is enormous and has important systemic liquidity implications. A survey  paper, reporting a dialogue on valuation under CSA and CCPs that could be a good informal introduction to this more technical paper is available in \cite{brigopallavicini2013dialogue}.

The context for this paper contribution is the following. Starting from summer 2007, with the spreading of the credit crunch, market quotes of forward rates and zero-coupon bonds began to violate standard no-arbitrage relationships. This was partly due to the liquidity crisis affecting credit lines, and to the possibility of a systemic break-down triggered by increased counterparty credit risk. Indeed, credit risk is only one facet of the problem, since the crisis started as a funding liquidity crisis, as shown for example by \cite{Tapking2009}, and it continued as a credit crisis following a typical spiral pattern as described in \cite{Brunnermeier2009}.

This has been the most dramatic signal for the inadequacy of standard financial modelling based on idealized assumptions on risk-free rates, idealized credit risk and on unrestricted access to funding instruments. Removing or relaxing such assumptions opens the door to financial models able to analyze the inner mechanics of a deal: collateral rules, funding policies, close-out procedures, market fees, among others. Such features require to be included in any realistic pricing framework and open a variety of quantitative and infrastructure challenges at an unprecedented level of difficulty.

A few papers in the financial literature have tried either to re-define the theory from scratch or to extend the standard framework to the new world. The works of \cite{Crepey2011,Crepey2012a,Crepey2012b} open the path towards a pricing theory based on different bank accounts, accruing at different rates, and representing the different cash sources one may have at her disposal to implement the hedging strategies, the funding policies and the margining rules. We cite also the partial works by \cite{Piterbarg2010,Piterbarg2012} (no default risk), focusing on perfectly collateralized deals and trying to reformulate the basic Black-Scholes theory, and the works of \cite{BurgardKjaer2011a,BurgardKjaer2011b} (no collateral). On the other hand, the works of \cite{Perini2011,Perini2012} stress the fact that all updated features can be included in terms of modified payoffs rather than through the need of a new and somehow ad-hoc pricing theory. In particular, the authors recognize that a financial contract must include hedging, funding and margining fees as specific and precisely defined additional cash flows. 

As a direct consequence of these approaches, the price of a derivative contract depends 
\begin{itemize}
\item on several choices the investor and the counterparty adopt for the collateralization procedure, either a bilateral Credit Support Annex (CSA) or the interposing action of a Central Clearing Counterparty (CCP), 
\item on the particular hedging strategy, which in turn may use collateralized instruments, 
\item and on the funding policy of the Treasury department that manages the collateral assets and the cash needed by the traders to fund their hedge. 
\end{itemize}

In more formal terms we can say that the pricing measure is now depending on the derivative contract itself, as shown in \cite{Perini2012}, and we lose the notion of a risk-free rate. A similar result can be found also in \cite{BieleckiRutkowski2013} in a more abstract framework. A full development and account of the related emerging pricing theory is given in the book \cite{BrigoMoriniPallavicini2012}.

Starting from such results \cite{BrigoPallavicini2013} discuss the consequences for pricing models. In particular, the authors focus on interest-rate models and on the pricing of interest-rate swaps. Under suitable assumptions they approximate the impact of hedging, funding and margining fees in pricing equations as a modification of discount factors and of forward rates. Furthermore, they stress the importance of including these corrections also in calibration or bootstrapping algorithms which are used to calculate the parameters of pricing models. As an example they re-formulate the multiple-curve HJM model of \cite{Moreni2010,Moreni2012} to be able to seize the impact in price of changing the collateralization procedure of a given interest-rate derivative. In particular, they discuss the case of interest-rate derivatives traded with a CCP extending the results of \cite{Rama2011}.

In this paper we try to distill the above contributions to highlight the common ingredients that a new pricing framework should include to be effective in a world where credit and liquidity issues cannot be disregarded. Furthermore, a key component of any quantitative analysis of trades is the margining procedure and the impact of initial and variation margins on the deal profitability.   We briefly summarize our earlier results in \cite{BrigoPallavicini2013} in order to be able to focus on concrete pricing cases such as partially collateralized deals, bilateral CSA deals, and contracts cleared through CCPs, the latter being quite sensitive following pressure from Dodd Frank and EMIR/CRD4 to move deals to CCPs. See for instance \cite{Arnsdorf2011}, \cite{Pirrong2011}, \cite{Rama2011}, and \cite{Heller2012}. We present both theoretical results and the discussion of numerical cases for the interest rate market. In particular, we describe how to price interest-rate swaps (IRS) cleared by a CCP or traded under a standard bilateral CSA with a particular attention to wrong-way risk and initial margins. We analyze the relative size of the different contributions due to credit and funding, wrong way risk, and draw conclusion on the overall patterns. A specific point that is made in this paper is that approaching an additive decomposition of funding, margining and credit costs is, in simplified contexts, possible. The fundamental nonlinearity of the valuation procedures in presence of funding costs has been pointed out earlier by us, among others, in \cite{Perini2011,Perini2012}. The funding rates that the treasury will apply depend on the future value of the same product we are pricing, and such value in turn depends on the future value of the funding rates. This leads to nonlinear and aggregation-dependent valuation that manifests itself in the nonlinearity of the pricing operator. While credit risk and CVA make the payoff nolinear, starting from a possibly linear one, funding costs have a more dramatic impact in that they make the pricing operator (or the related expectation) nonlinear. A consequence of this is that the partition of valuation adjustments in purely credit and funding ones is not possible in general. However, in this paper, through a number of specific simplifying choices with precise financial interpretation, we achieve a quasi separable decomposition of the total valuation adjustment into funding, credit and margining costs and illustrate it into the above mentioned numerical examples.

The paper is structured as follows.

In Section \ref{sec:pricing} we introduce CCPs and CSA trades in detail, looking both at the systemic/macro implications and at the micro/single player view. Variation and initial margins, in particular, are introduced and discussed.

In Section \ref{sec:newpf} we summarize the valuation theory developed previously in \cite{Perini2011,Perini2012,BrigoPallavicini2013}, see also the non-technical survey paper
\cite{brigopallavicini2013dialogue}. We also update the previous theory to include Initial Margins specifically and CCP mechanics, and possibly delays in the closeout procedure. This is an important feature that was missing in our previous works, so that this paper presents some novelty also in the general methodology. 

In Section \ref{sec:IRD} we apply the general valuation apparatus to interest rate modeling. By specific assumptions on the exposure, on the information (filtrations), on contagion and on margining we achieve a way to contain nonlinearities in the pricing operator and to obtain a decomposition into Credit, Debit, Funding and Margining Valuation Adjustments (CVA, DVA, FVA, MVA) over a basic mark to market. 

In Section \ref{sec:numerics} we apply the methodology to a numerical case study and illustrate the relative size of the different adjustments and the wrong way risk patterns through the study of an interest rate swap portfolio.

Section \ref{sec:conclusion} concludes the paper.

\section{Valuation under Credit, Collateral and Funding: CCPs and CSAs}
\label{sec:pricing}

After the onset of the crisis in 2007, under the pressure of the liquidity crisis reducing the credit lines along with the fear of an imminent systemic break-down, all market instruments began to be quoted by taking into account, more or less implicitly, credit and funding liquidity adjustments. As a consequence classical pricing models, which often ignore credit and liquidity issues, require a careful check of standard theoretical assumptions. Assumptions and approximations stemming from abstract pricing theory should be replaced by strategies implemented with market instruments. We need to go back to market observables by limiting ourselves to price only derivative contracts we are able to replicate by means of market instruments. The comparison to market data and processes is the only means we have to validate our theoretical assumptions, so as to drop them if in contrast with observations.

A paradigmatic example is the inclusion of credit valuation adjustments (CVA) for interest-rate instruments, such as those analyzed in \cite{BrigoPallavicini2007}, that breaks the relationship between risk-free zero coupon bonds and LIBOR forward rates. Also, funding in domestic currency on different time horizons must include counterparty risk adjustments and liquidity issues, see \cite{Filipovic2012}, breaking again this relationship. A direct consequence is the impossibility to describe all LIBOR rates in terms of a unique zero-coupon yield curve. Indeed, since 2009 and even earlier, we had evidence that the money market for the Euro area was moving to a multi-curve setting. Many papers were devoted to the topics, by starting from the first papers by \cite {Henrard2007,Henrard2009}, \cite{Kijima2009}, \cite{Mercurio2009,Mercurio2010}, and \cite{PallaviciniTarenghi}.

Moreover, the growing attention on counterparty credit risk is transforming OTC derivatives money markets. An increasing number of derivative contracts is cleared by CCPs, while most of the remaining contracts are traded under collateralization, regulated by a Credit Support Annex (CSA). Both cleared and CSA deals require collateral posting, as default insurance, along with its remuneration. We cannot neglect such effects on valuation. 

A general point on our terminology is that whenever we use the term "bilateral trade" in a context with collateral, we are referring to a deal covered by a bilateral CSA and not to a deal going through a CCP. The short sentence "cleared trade" refers instead to a deal going through a CCP. Part of our analysis applies also to bilateral trades without collateral at all, although our main interest in the present paper is the margining process.

In the following sections we look in detail to derivative contracts to list all the cash flows and fees needed to effectively close a position. Even if different asset classes have derivatives markets that may operate differently, the general principles of our approach are not linked to a particular asset class. Nonetheless, in this paper we focus, as a main example, on interest-rate derivatives, which are a representative case given interest rate swaps prevalence in cleared contracts. In particular, we analyze the collateral procedure and the hedging strategy to detect the presence of possible funding costs. Before looking at this in detail, we introduce CCPs and margining more in detail.

\subsection{Valuation in presence of CCPs: a few basic questions}

A possible attitude to the onset of collateralization, either through CCPs or through bilateral collateralized trades (under CSA), is that this will make all calculations of counterparty risk pricing and/or funding costs either trivial or irrelevant. For example, CVA could disappear, while Funding Valuation Adjustments would be trivially covered by margining costs. The methodological complexity highlighted in earlier works such as for example \cite{Perini2012}, \cite{BrigoMoriniPallavicini2012}, and \cite{BrigoPallavicini2013} would become irrelevant. We believe this is a naive stance and devote now a short section to criticize this conception of valuation, hedging, credit and liquidity risk and collateral modeling. 

Let us start with an obvious question: what are CCPs and what is their role? CCPs are commercial entities that, ideally, would interpose themselves between the two parties in a trade. 

\begin{figure}
\begin{center}
\scalebox{1}{
\begin{tikzpicture}
   \node [trader, node distance=4cm] (D1) {Client};
   \node [trader, right of=D1, node distance=4cm] (D2) {Client};
   \path [to,transform canvas={yshift=3pt}] (D1) -- node [above] {IM} (D2);
   \path [from,transform canvas={yshift=6pt}] (D1) -- node [above] {} (D2);
   \path [to,transform canvas={yshift=-3pt}] (D1) -- node [below] {} (D2);
   \path [from,transform canvas={yshift=-6pt}] (D1) -- node [below] {VM} (D2);
\end{tikzpicture}}
\end{center}
\caption{Flows of initial and variation margin within counterparties acting in a bilateral contract. IM stands for initial margin, while VM for variation margin. The two margins flow in both directions. }
\label{fig:bilateral}
\end{figure}
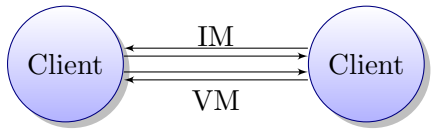

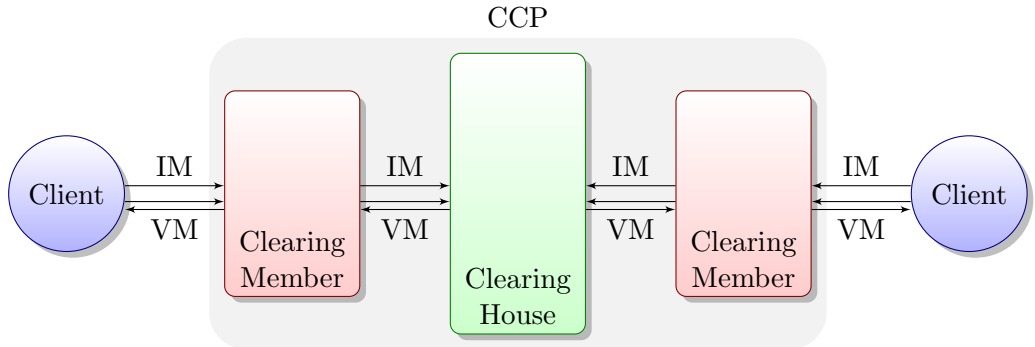
\begin{figure}
\begin{center}
\scalebox{1}{
\begin{tikzpicture}
   \node [trader, node distance=3cm] (D1) {Client};
   \node [bank, right of=D1, text height=2cm,node distance = 3cm] (CM1) {Clearing Member};
   \node [market, right of=CM1, text height=3cm, node distance = 3cm] (CCP) {Clearing House};
   \node [bank, right of=CCP, text height=2cm,node distance = 3cm] (CM2) {Clearing Member};
   \node [trader, right of=CM2, node distance=3cm] (D2) {Client};
   \path [to,transform canvas={yshift=3pt}] (D1) -- node [above] {IM} (CM1);
   \path [to,transform canvas={yshift=-3pt}] (D1) -- node [below] {} (CM1);
   \path [from,transform canvas={yshift=-6pt}] (D1) -- node [below] {VM} (CM1);
   \path [to,transform canvas={yshift=3pt}] (CM1) -- node [above] {IM} (CCP);
   \path [to,transform canvas={yshift=-3pt}] (CM1) -- node [below] {} (CCP);
   \path [from,transform canvas={yshift=-6pt}] (CM1) -- node [below] {VM} (CCP);
   \path [from,transform canvas={yshift=3pt}] (CM2) -- node [above] {IM} (D2);
   \path [to,transform canvas={yshift=-3pt}] (D2) -- node [below] {} (CM2);
   \path [from,transform canvas={yshift=-6pt}] (D2) -- node [below] {VM} (CM2);
   \path [from,transform canvas={yshift=3pt}] (CCP) -- node [above] {IM} (CM2);
   \path [to,transform canvas={yshift=-3pt}] (CM2) -- node [below] {} (CCP);
   \path [from,transform canvas={yshift=-6pt}] (CM2) -- node [below] {VM} (CCP);
   \begin{pgfonlayer}{background}
   \node [background, fit=(CM1) (CM2) (CCP), label=above:CCP] {};
   \end{pgfonlayer}
\end{tikzpicture}}
\end{center}
\caption{Flows of initial and variation margin within counterparties acting in a centrally cleared contract. IM stands for initial margin, and it flows only towards the CCP. VM stands for variation margin, and it flows in both directions. }
\label{fig:clearing}
\end{figure}

More specifically, a CCP acts as a market participant who is taking the risk of the counterparty default and ensures that the payments are performed even in case of default. To achieve this an initial bilateral trade, shown in figure \ref{fig:bilateral}, is split into two trades, with the CCP standing in between the parties (clients). In practice, the counterparties operate with the CCP by means of intermediate clearing members as in the scheme listed in figure \ref{fig:clearing}.

\begin{itemize}
\item Each client enters into a deal with his clearing member which offsets the position with the CCP. There are no more direct obligations between the two clients.
\item Each party will post collateral margins say daily, every time the mark-to-market goes against that party. This is called Variation Margin (VM). 
\item Moreover, there is also an initial margin (IM) that is supposed to cover for additional risks like deteriorating quality of collateral, gap risk, wrong way risk, and so on. This may be posted at need. 
\item VM posted by one party is passed by the CCP to the other party in the trade, whereas IM will be held by the CCP and will not be passed to the other party.  
\item If a party in the deal defaults and the mark-to-market is in favour of the other party, then the surviving party will keep the VM and also possibly obtain the IM  from the CCP and will not be affected, in principle, by counterparty risk. 
\end{itemize}

CCPs will reduce risk in many cases but are not a panacea. They require daily margining, and one may question: (i) the pricing of the fees they apply, (ii) the appropriateness of the initial margins and of over-collateralization buffers that are supposed to account for wrong-way risk and collateral gap-risk, among other features, and (iii) the default risk of Clearing Members and, even, of CCPs themselves.

An important issue for clients is that the CCP does not post part of the collateral (initial margins) directly to the entities trading with it, as the collateral agreement is not symmetric. In a scenario of a CCP default one has the over-collateralization cost to lose in a uncollateralized type CVA situation. Hopefully, the default probability is low, making CVA small, bar strong contagion, gap risk and wrong-way risk, that may or may not be adequately covered by initial margins. However unlikely, however, CCPs defaults have not been uncommon in the past.
Defaulted CCPs include the following. 1974: Caisse de Liquidation des Affaires en Marchandises; 1983: Kuala Lumpur Commodity Clearing House; 1987: Hong Kong Futures Exchange. The ones that were close to default are 1987: CME  and OCC, USA; 1999: BM{\&}F, Brazil.

Valuation of the above risks requires CVA and funding costs type analytics, inclusive of collateral gap-risk and wrong-way risk, similar to those discussed in earlier papers and books, see for example \cite{Perini2012},  \cite{BrigoMoriniPallavicini2012} and  \cite{BrigoPallavicini2013}. Indeed, unless one trusts blindly a specific clearing house, it will be still necessary to access residual CVA and funding analytics and risk measures. There are also important considerations on the macroeconomic and systemic effects of CCPs, and while we do not discuss such aspects in details here, to those we turn now briefly. As discussed in \cite{Piron2012}, the above CCPs default are to be kept in mind, and further one needs to note that CCPs are usually highly capitalized. All clearing members post collateral in an asymmetric way. Initial margin means clearing members are over-collateralized all the time. The TABB Group says extra collateral could be about 2 $\$$ Trillion \cite{Tabb2011}.

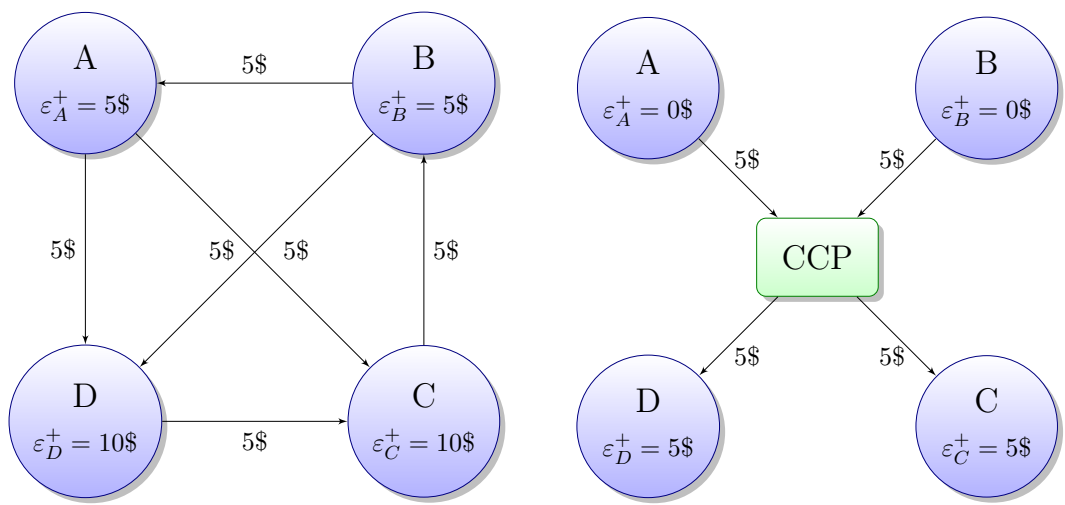
\begin{figure}
\begin{center}
\scalebox{0.9}{
\begin{tikzpicture}
   \node [trader, node distance=5cm] (A) {\shortstack[cc]{{\Large A}\medskip\\$\varepsilon_A^+=5\$$}};
   \node [trader, right of=A, node distance=5cm] (B) {\shortstack[cc]{{\Large B}\medskip\\$\varepsilon_B^+=5\$$}};
   \node [trader, below of=B, node distance=5cm] (C) {\shortstack[cc]{{\Large C}\medskip\\$\varepsilon_C^+=10\$$}};
   \node [trader, below of=A, node distance=5cm] (D) {\shortstack[cc]{{\Large D}\medskip\\$\varepsilon_D^+=10\$$}};
   \path [from] (A) -- node [above] {$5\$$} (B);
   \path [to] (A) -- node [left] {$5\$$~~} (C);
   \path [to] (A) -- node [left] {$5\$$} (D);
   \path [from] (B) -- node [right] {$5\$$} (C);
   \path [to] (B) -- node [right] {~~$5\$$} (D);
   \path [from] (C) -- node [below] {$5\$$} (D);
\end{tikzpicture}}~~~~~~
\scalebox{0.9}{
\begin{tikzpicture}
   \node [trader, node distance=5cm] (A) {\shortstack[cc]{{\Large A}\medskip\\$\varepsilon^+_A=0\$$}};
   \node [trader, right of=A, node distance=5cm] (B) {\shortstack[cc]{{\Large B}\medskip\\$\varepsilon^+_B=0\$$}};
   \node [trader, below of=B, node distance=5cm] (C) {\shortstack[cc]{{\Large C}\medskip\\$\varepsilon^+_C=5\$$}};
   \node [trader, below of=A, node distance=5cm] (D) {\shortstack[cc]{{\Large D}\medskip\\$\varepsilon^+_D=5\$$}};
   \node [market, below right of=A, node distance=3.535cm] (X) {{\Large CCP}};
   \path [to] (A) -- node [above] {~~$5\$$} (X);
   \path [to] (B) -- node [above] {$5\$$~~} (X);
   \path [from] (C) -- node [below] {$5\$$~~} (X);
   \path [from] (D) -- node [below] {~~$5\$$} (X);
\end{tikzpicture}}
\end{center}
\caption{Bilateral trades and exposures without CCPs (on the left) and with CPPs (on the right). Each node lists the sum of positive exposures, each arrow the due cash flows. The diagram refers to the discussion in \cite{Kiff2009}.\label{fig:kiff}}
\end{figure}

On the positive side, there is undoubtedly an important netting benefit with CCPs, and this is illustrated with a toy example in Figure \ref{fig:kiff}.

However, in reality this is not so clear. A typical bank may have a quite large number of outstanding trades, making the netting clause quite material. With just one CCP for all asset classes across countries and continents, netting efficiency would certainly improve. However, in real life CCPs deal with specific asset classes or geographical areas, and this may even reduce netting efficiency compared to now, as discussed in \cite{Rama2013}, that refines the earlier analysis in \cite{duffiezhu2011}.

CCPs compete with each other and one can be competitive in specific areas but hardly in all of them. Some CCPs will be profitable in specific asset classes and countries. They will deal mostly with standardized transactions. Even if CCPs could function across countries, bankruptcy laws can make collateral held in one place unusable to cover losses in other places, as pointed out in \cite{Singh2011b}.

The geographical angle seems to be an issue, with no international law addressing how CCPs would connect through EMIR/CDR 4/Basel III and DFA. For example, as of 2012 there is ``No legal construct to satisfy both Dodd Frank Act and EMIR and allow EU clients to access non-EU CCPs' \cite{Wayne2012}. There are also other conflicts in this respect. Where will CCPs be located and which countries will they serve? For example, the European Central Bank is adopting a rule that limits operations of UK based CCPs such as LCH--Clearnet for Euro denominated deals. This lead to a legal battle with the UK government invoking the European Court of Justice. 

To compete CCPs may lower margin requirements, which would make them riskier, remembering also the above CCPs defaults. In the US, where the OTC derivatives market is going through slightly more than 10 large dealers and is largely concentrated among 5, we could have a conflict of interest. If CCPs end up incorporating most trades currently occurring OTC bilaterally, then CCPs could become ``too big to fail'', \cite{Miller2011}.

We may conclude this introduction by saying that even with CCPs in place, one needs a strong analytical and numerical apparatus for pricing/hedging and risk. For all the reasons we illustrated, CCPs are not the end of CVA and its funding costs extensions, and this paper is devoted precisely at trades valuation in presence of CCPs or CSAs, or simply of bilateral trades under credit and funding costs.  Our  general framework takes into account residual credit risk and gap risk, initial and variation margins, collateral, close-out netting rules, and wrong-way risk. Our approach is fully arbitrage free but based only on market observables. We will develop this theory and its applications formally in Section \ref{sec:newpf}, seeing that basic financial facts lead to complex mathematical tools such as BSDEs. However, before looking at that, we introduce a more detailed analysis of CCPs and CSAs, and further we introduce basic formulas for including initial and variation margins into valuation. 

\subsection{Trading via CSA or CCP: Analogies and Differences}
\label{sec:ccp}

Since many years clearing houses (CCP) started to clear interest-rate swaps. For instance, LCH after a few years of studies introduced in 1999 clearing services for plain-vanilla interest-rate swaps in four major currencies. After the crisis of 2007 an increasing number of operations moved from a bilateral OTC agreement under CSA to a cleared trade. As we have seen above, a CCP interposes itself between the buyer and the seller of trades, becoming their new counterparty. In order to protect itself from market and credit risks, the Clearing House calculates exposures and calls margin daily. If one decides not to trade through a CCP, one may still decide to exchange collateral margins daily with the counterparty in a more private setting. With an eye to this situation, in 2011 ISDA developed a proposal for CSA agreements which is in accordance with the collateralization practices adopted by clearing houses, and then issued in 2013 as standard CSA (SCSA), see \cite{ISDA2013}. The SCSA aims at a similar treatment of collateralization for bilateral and cleared trades: it restricts eligible collateral for variation margins to cash, promotes the adoption of overnight rate as collateral rate, and factors in currency risks for the collateral calculation. 

In a cleared trade the counterparty of a CCP has to insure the clearing house against any credit or market risk by posting collateral, known as  variation margin, to match the mark-to-market variation of the deal. It is market practice to pay interest on cash collateral. CCPs usually compensate variation margin with overnight interest rates, as done in identical uncleared bilateral trades under CSA. The analysis of the impact of variation margin procedures can be found in \cite{Rama2011}, where convexity adjustments and NPV effects are discussed for different clearing houses. These two effects are also discussed in \cite{BrigoPallavicini2013}.

Moreover, the counterparty has to cover any potential future exposures resulting in replacement costs, as discussed in \cite{Heller2012}. In particular, a CCP is vulnerable to losses on defaulting counterparty exposures between the time of the last variation margin payment of the defaulting counterparty and close-out valuation. This is known as margin period of risk. Furthermore, initial margin may cover gap risks, namely adverse variations in mark to market since the last collateral posting. Initial margins are a source of funding costs for the counterparty, which is forced to post additional collateral assets to protect the CCP. On the other hand, initial margins are not posted by the CCP to the counterparty, so that they do not result in a funding benefit. 

In the case of a CSA, independent amounts are posted at the beginning of the trade, to cover for additional risks that are not met by variation margins, for example gap risk. However, with the standard CSA we are moving towards an Initial Margins approach similar to the approach of CCPs, where the initial margin may be paid several times during the life of the deal in case there is a perceived deterioration of gap risk and related issues.

\subsection{CCP: Clearing Members and Recoveries}
\label{sec:mechanics}

We look now at the trade in place between one client and his clearing member, and, in particular, at the collateral procedure. We identify the client with the letter ``C'' used to name the counterparty of the deal in section \ref{sec:newpf} below, and the clearing member with the letter ``I'' used to name the investment bank below.

The two parties post and receive the variation margin, while only the client posts the initial margin. The variation margin can be re-hypothecated, while the initial margin is segregated in a different account. In case of re-hypothecation, the collateral provider must therefore consider the possibility to recover only a fraction of his collateral. If the clearing member is the collateral taker, we denote the recovery fraction on re-hypothecated collateral by $\rec'_I$, while if the client is the collateral taker, then we denote the recovery fraction on re-hypothecated collateral by $\rec'_C$. Accordingly, we define the collateral loss incurred by the client upon clearing member default by $\lgd'_I = 1 - \rec'_I$ and the collateral loss incurred by the clearing member upon client default by $\lgd'_C = 1 - \rec'_C$.

In a CSA bilateral deal, typically, the surviving party has precedence on other creditors to get back his collateral, thus $\rec_I \leq \rec_I' \leq 1$, and $\rec_C \leq \rec_C' \leq 1$. Here, $\rec_I$ ($\rec_C$) denote the recovery fraction of the market value of the transaction that the client (clearing member) gets when the clearing member (client) defaults.

On the other hand, in the specific case of a deal cleared via a CCP, we have that, in case of default of the clearing member, the deal is transferred to backup clearing members (if any), so that recovery rates should be equal to one. Yet, the transfer procedure is implemented in terms of a competitive auction, where the backup clearing members are not forced to accept the client position, which, if rejected by them, is liquidated on the market. As a result losses may occur from this procedure, leading to recovery fractions possibly less than one.

In the following sections we describe the mechanics of centrally cleared contracts from the point of view of the client of a clearing broker, since we aim at evaluating its replication price for a cleared contract. Furthermore, we will discuss also the case of a bilateral contract when the initial margin is posted. 

\subsection{Funding Netting Sets, Margins and Re-hypothecation}
\label{sec:nettingset}

The different prescriptions on variation and initial margins re-hypothecation must be considered when evaluating the cash needed to implement the hedging strategy, which, in turn, affects evaluation of funding costs (we will give a precise expression for these costs in equations \eqref{eq:varphi_bilateral} and \eqref{eq:varphi_ccp} below). Here, we generalize the discussion of \cite{Perini2011} to a collateralization strategy based on variation and initial margins.

In February 2013 the Basel Committee on Banking Supervision (BCBS) and the International Organization of Securities Commissions (IOSCO) issued a second consultative document on margin requirements for non-centrally cleared derivatives \cite{BaselIM}. The aim is to introduce minimum standards for initial margin posting for non-centrally cleared derivatives. The document discusses the methodologies for calculating initial and variation margins in OTC derivatives traded between financial firms and systemically-important non-financial entities. The principles guiding the proposal promote a margining practice similar to the one adopted for centrally cleared products. \emph{This is a very important point in that CSA bilateral deals will resemble more and more CCP cleared trades, with Initial and Variation Margins features. In particularly, the newly proposed Initial Margin goes beyond the previous notion of Independent Amounts.}

According to the document we should introduce two initial margin accounts, one for each party of the deal, so that we define three adapted processes $M_t$, $N^C_t$ and $N^I_t$ to model respectively the variation margin, the initial margin posted by the counterparty, and the initial margin posted by the investor, which represent the collateral assets exchanged by the margining procedures, so that the total amount $C_t$ of collateral assets, from the point of view of ``I'', is given by 
\begin{equation}
\label{eq:collateral}
C_t := M_t + N^C_t + N^I_t
\;,\quad
N^C_t \geq 0
\;,\quad
N^I_t \leq 0 \,.
\end{equation}%
Each party is posting the initial margin without netting between the two accounts to protect the other party from gap risk.

The case of a contract cleared by a CCP can be simply obtained by setting $N^I_t = 0$, since in this case the client "C" is posting the initial margin, whereas the clearing member "I" is not.
We now turn to re-hypothecation.

Different re-hypothecation clauses can hold on the variation and the initial margin accounts. As a first possibility we assume that the variation margin can be re-hypothecated, so that we can use it to reduce the cash needed by the hedging strategy, namely we can consider the derivative and the variation margin as part of the same funding netting set. Thus, we can replicate the derivative price by means of the following hedge
\begin{equation}
\label{eq:replica}
V_t = F_t + M_t + H_t \,.
\end{equation}%
Here the value $V_t$ of the claim is composed by the cash part $F_t$ of the hedge portfolio, by the collateral part given by the variation margin $M_t$, and by the risky component of the hedge portfolio, namely $H_t$.

To fix ideas, let us think of an equity call option contract in the standard Black and Scholes setting. Let $S_t$ denote the price process of the underlying risky equity asset, and assume constant and deterministic interest rates $r$. The risky hedge portfolio is the delta position in the equity, $H_t = \Phi(d_1(t,S_t)) S_t$, whereas the cash part of the hedge is the bank account position needed to complete the hedge, namely $ F_t = - K e^{-r (T-t)}\Phi(d_2(t,S_t))$, and finally there is no margin, $M=0$. Here $\Phi$ is the cumulative distribution function of the standard normal distribution and the $d_1$ and $d_2$ terms are the familiar Black-Scholes formula terms that may be found in any option pricing textbook. 

Going back to the general theory, we can extend our argument at portfolio level by adding up all variation margins (when re-hypothecation is allowed) coming from derivatives belonging to the same funding netting set. On the other hand, if the variation margin cannot be re-hypothecated we have
\[
V_t = F_t + H_t \,.
\]%
In the following we assume that the variation margin can always be re-hypothecated. Formulas for the opposite case follow in a straightforward manner.

It is now time to finally introduce the technicalities needed to face pricing in presence of CCPs or CSAs. As mentioned earlier, this entails modeling credit risk and funding cost, the margining process and the close-out netting rules, all in terms of precise cash flows and in line with arbitrage-free derivatives valuation. This is where the technical part of the paper begins.

\section{A New Pricing Framework}
\label{sec:newpf}

In order to price a financial product (for example a derivative contract), we have to consider all the cash flows occurring after the trading position is entered. We can group them as follows:
\begin{enumerate}
\item product cash flows (e.g. coupons, dividends, premiums, final payout, etc.) inclusive of hedging instruments cash flows;
\item cash flows required by the collateral margining procedure;
\item cash flows required by the funding and investing (borrowing and lending) procedures;
\item cash flows occurring upon default (close-out procedure).
\end{enumerate}%

We refer to the two names involved in the financial transaction and subject to default risk as investor (also called name ``I'', usually the bank, or the clearing member of CCP) and counterparty (also called name ``C'', for example a corporate client, but also another bank, see again Figures \ref{fig:bilateral} and \ref{fig:clearing}). 

We denote by $\tau_I$ and $\tau_C$ respectively the default times of the investor and counterparty. We fix the portfolio time horizon $T>0$, and fix the risk-neutral pricing model $(\Omega,\mathcal{G},\mathbb{Q})$, with a filtration $(\mathcal{G}_t)_{t \in [0,T]}$ such that $\tau_C$, $\tau_I$ are $(\mathcal{G}_t)_{t \in [0,T]}$-stopping times. We denote by $\ExG{t}{\cdot}$ the conditional expectation under $\mathbb{Q}$ given $\mathcal{G}_t$, and by $\ExG{\tau_i}{\cdot}$ the conditional expectation under $\mathbb{Q}$ given the stopped filtration $\mathcal{G}_{\tau_i}$. We exclude the possibility of simultaneous defaults, and define the first default event between the two parties as the stopping time
\[
\tau := \tau_C \wedge \tau_I \,.
\]
We will also consider the default-free market sub-filtration $({\cal F}_t)_{t \ge 0}$ that one obtains implicitly by assuming a separable structure for the complete market filtration $({\cal G}_t)_{t \ge 0}$. ${\cal G}_t$ is then generated by the pure default-free market filtration ${\cal F}_t$ and by the filtration generated by all the relevant default times monitored up to $t$ (see for example \cite{BieleckiRutkowski2002}). 

\subsection{The Master Pricing Equation in Presence of Credit, Collateral and Funding}

The price of a derivative, inclusive of collateralized credit and debit risk (CVA and DVA), margining costs, and funding and investing costs can be derived by following \cite{Perini2011,Perini2012}, and is given by the following master equation:
\begin{equation}
\label{eq:fundingpreview}
V_t = \ExG{t}{\Pi(t,T\wedge\tau) + \gamma(t,T\wedge\tau) + \varphi(t,T\wedge\tau) + \ind{t<\tau<T} D(t,\tau) \theta_{\tau} } \,,
\end{equation}%
where the terms are defined below\footnote{In \cite{Perini2011,Perini2012} the derivative price inclusive of credit an funding effects is denoted ${\bar V}_t$ and the market rates are marked with a tilde. This is done to distinguish them from funding costs-free prices.  Here, we use $V_t$ for the full price and we drop the tilde for notation convenience.}.
\begin{itemize}

\item Notation $D(t,u)$ in general will denote the risk-neutral default-free and funding-free discount factor, given by the ratio 
\[
D(t,u) = B_t/B_u
\;,\quad
d B_t = r_t B_t dt \,,
\]%
where $B$ is the bank account numeraire, driven by the risk free instantaneous interest rate $r$ and associated to the risk-neutral measure $\mathbb{Q}$.

\item $\Pi(t,T)$ is the sum of all discounted payoff terms in the interval $(t,T]$, without credit or debit risk, without funding costs and without collateral cash flows. In other terms, these are the financial instrument discounted cash flows without additional risks.

\item $\gamma(t,T)$ represents the collateral margining costs within the interval $(t,T]$, and they are equal to the cost-of-carry of the collateral account $C_t$, namely to
\begin{equation}
\label{eq:gamma}
\gamma(t,u) := \int_t^u dv \, ( r_v - c_v ) C_v D(t,v) \,,
\end{equation}%
where the collateral rate $c_t$ is defined by the CSA holding between the two counterparties, or by the CCP contractual rules.

\item $\varphi(t,T)$ represents the funding and investing costs within the interval $(t,T]$, which consists in the fees payed to the Treasury department to obtain the cash (or the benefits received when the cash is to be invested), plus the spread over Treasury rates to be payed (or received) to access the market of risky assets. We will explicitly define such costs for bilateral and CCP-cleared trades in the following section where we extend the analysis of \cite{Perini2012}.

\item $\theta_{\tau}$ is the price at $\tau$ of the cash flow payed or received at the end of the default procedure. The cash flow depends on the value of the collateral account and on the residual value of the claim being traded at default, also interpreted as the replacement cost for the deal at default (close-out amount). It is primarily this term that originates the familiar Credit and Debit Valuation Adjustments (CVA/DVA) terms. In the following we generalize the analysis of \cite{BrigoPallaviciniPapatheodorou} to apply it both in presence of an explicit margin period of risk and when many margin accounts are present.

\end{itemize}

The above definitions link the master pricing equation \eqref{eq:fundingpreview} to the collateralization procedure holding between the two counterparties, and to the funding policies adopted by them. As a direct consequence the replication price of contract, holding between two counterparties under a bilateral CSA, is different from the price of the same contact when cleared with a CCP. The same happens if the Treasury adopts different funding policies or the collateral accounts have to be segregated not contributing to the funding netting set of the derivative. In this paper we try to describe such different scenarios by illustrating the main ingredients we have to take into account when pricing bilateral or centrally cleared contracts to gauge the impact of credit risk and funding costs. We will make reasonable assumptions on contractual issues when the pricing framework does not force us to a particular a choice with the aim to present numerical results.

Now that we have introduced the pricing framework in presence of credit, collateral, margining and funding, including close-out netting rules and CVA type adjustments, we may turn to applying this framework to pricing deals when a CCP is interposed or when a bilateral CSA is present between two trading entities (e.g. the recent SCSA).

\subsection{Funding Costs under Bilateral CSA or CCP Clearing}

We recall that we value the contract from the point of view of the investor ``I''.

First, we start by presenting the definition of funding costs given by \cite{Perini2012}, which applies to the case of a bilateral contract with collateral re-hypothecation, namely to the case $N_t=0$. We have
\begin{equation}
\label{eq:varphi_bilateral}
\varphi(t,u) := \int_t^u dv \, ( r_v - f_v ) F_v D(t,v) - \int_t^u dv \, ( f_v - h_v ) H_v D(t,v) \;,\quad N_t \doteq 0
\end{equation}%
where $f_t$ is the Treasury rate that is established according to the funding policy, while $h_t$ is the rate charged by the market to trade the hedging instruments contained in the account $H_t$. The accounts $F_t$ and $H_t$ contain respectively the cash and the risky assets positions needed in the hedging strategy, see our earlier example in the Black and Scholes setting. We will define these rates $f_t$ and $h_t$ in the following, according to the specific pricing problem.

It is worth clarifying in detail where the different terms in the equation for $\varphi$ come from. The amount $-(f- r) F$ is interpreted as the cost of funding for the cash part of the hedge. Indeed, if we are for example in a position where $F > 0$ (and, typically, assume $f > r$) we are borrowing cash from our treasury to keep the hedge $F$ going. This cash is lent to our trading desk and we have to pay (hence the minus sign) the treasury an amount of interest given by the spread of the funding rate $f$ over the risk free rate $r$. In the other case, if $F$ is negative, we have a positive amount $(f - r)(- F)$. Indeed, in this case the treasury will pay us interest for the cash we are shorting, and this interest will be the treasury rate minus the risk free rate, which is for us a positive funding benefit rather than a cost. 

As for the second term, $-(f-h) H$, it is best to think this term as $-(f-r) H + (h-r) H$. Assuming the hedge $H$ to be positive, we need to borrow risky asset to maintain it. To get this risky asset we need an amount of cash $H$ equal to the asset price, and on this cash, which we borrow from the treasury, we pay (hence the minus sign) the usual interest $f-r$. On the other hand, as soon as we have the risky asset, we can repo-lend it at a rate $h$, gaining (hence the plus sign) a spread $h-r$. 


The above definition is valid if all the accounts which need funding are included within the funding netting set defining the account $F_t$. If additional accounts are needed to implement the collateralization procedure, for example segregated initial margins, as with CCP cleared contracts and bilateral contracts under CSA with the forthcoming rules, then their funding costs must be added to equation \eqref{eq:varphi_bilateral}. If we consider additional initial margins, which are kept into a segregated account, one posted by the investor ($N^I_t\leq0$) and one by the counterparty ($N^C_t\geq0$), we obtain
\begin{eqnarray}
\label{eq:varphi_ccp}
\varphi(t,u) &:=& \int_t^u dv \, ( r_v - f_v ) F_v D(t,v) - \int_t^u dv \, ( f_v - h_v ) H_v D(t,v) \\\nonumber
             & +& \int_t^u dv ( f^{N^C}_v - r_v ) N^C_v + \int_t^u dv ( f^{N^I}_v - r_v ) N^I_v \,,
\end{eqnarray}%
where $f^{N^C}_t$ and $f^{N^I}_t$ are the funding rate assigned by the Treasury department to the initial margin account, which, in line of principle, could be different from $f_t$ since the initial margins are not in the funding netting set of the derivative.

The last two terms can be understood as follows. Consider, for instance, that the funding rate is greater than the risk-free rate. The party which is posting the initial margin has a penalty given by the cost of funding this extra collateral, while the party which is receiving it reports a funding benefit, but only if the contractual rules allow him to invest the collateral in some low-risk activity, otherwise the funding rate for the initial margin will be simply equal to the risk free rate, and there are no price adjustments.

If we put this result back into the pricing equation \eqref{eq:fundingpreview}, we get
\begin{eqnarray}
\label{eq:fundingwithnettingset}
V_t
& = & \int_t^T \,\ExG{t}{ \ind{u<\tau} D(t,u) ( \Pi(u,u+du) + \ind{\tau\in du} \theta_u ) } \\\nonumber
& + & \int_t^T du \,\ExG{t}{ \ind{u<\tau} D(t,u) ( ( f_u - c_u ) M_u + ( r_u - f_u ) V_u - ( r_u - h_u ) H_u ) } \\\nonumber
& + & \int_t^T du \,\ExG{t}{ \ind{u<\tau} D(t,u) ( ( f^{N^C}_u - c_u ) N^C_u + ( f^{N^I}_u - c_u ) N^I_u ) }
\end{eqnarray}%
where we have expressed as an integral also the first two terms. The above formula is not a closed form formula, but an equation. This is because $V_t$ appears also on the right hand side and, furthermore, the rates on the right hand side might depend on $V_t$ itself, as we will see explicitly in Section \ref{sec:hifc} below. A formal application of the Feynman-Kac formula allows one to obtain a nonlinear PDE formulation for this pricing problem, see for example \cite{Perini2011,Perini2012}. We now rewrite the above equation as a backward stochastic differential equation (BSDE). 

\subsection{BSDE Formulation}
\label{sec:bsde}

The pricing equation can be reformulated as a backward stochastic differential equation (BSDE) with terminal condition at the minimum time between default time and maturity. A similar formulation in the case of bilateral trades can be found in \cite{Crepey2011}. We multiply both sides of \eqref{eq:fundingwithnettingset} by $D(0,t)$ and then we add
\[
X_t := \int_0^t du\, \ind{u<\tau} D(0,u) \left( \Pi(u,u+du) + \ind{\tau\in du} \theta_u +  ( f_u - c_u ) M_u \right.\]\[  \left. + ( f^{N^C}_u - c_u ) N^C_u + f^{N^I}_u - c_u ) N^I_u + ( r_u - f_u ) V_u - ( r_u - h_u ) H_u \right)
\]
on both sides. We obtain
\[
D(0,t) V_t + X_t = \ExG{t}{X_T} \,.
\]
The right hand side is clearly a ${\cal G}_t$ martingale under the risk neutral measure, call it ${\cal M}_t$. Now differentiate with respect to time $t$ both sides to obtain
\[
- r_t D(0,t) V_t dt +  D(0,t) dV_t + d X_t =  d {\cal M}_t \,.
\]
Substituting for $dX_t$ leads to the BSDE formulation for $t\in[0,\tau\wedge T]$:
\begin{equation}
\label{eq:bsde_rn}
dV_t - f_t V_t \,dt + \Pi(t,t+dt) + ( f_t - c_t ) M_t \,dt + ( f^{N^C}_t - c_t ) N^C_t \,dt + ( f^{N^I}_t - c_t ) N^I_t \,dt - ( r_t - h_t ) H_t \,dt = d {\cal M}_t
\end{equation}%
where the terminal condition is set at $\tau\wedge T$ as
\begin{equation}
\label{eq:bsde_tc}
V_{\tau\wedge T} = \ind{\tau<T} \theta_{\tau} \,.
\end{equation}%

If we assume that the underlying risk factors are described by a Markov diffusion, we can apply the It\^o formula to obtain the following BSDE formulation which does not depend any longer on the risk-free rate
\begin{equation}
\label{eq:bsde}
dV_t - f_t V_t \,dt + \Pi(t,t+dt) + ( f_t - c_t ) M_t \,dt + ( f^{N^C}_t - c_t ) N^C_t \,dt + ( f^{N^I}_t - c_t ) N^I_t \,dt = d {\cal M}^h_t
\end{equation}%
where ${\cal M}^h_t$ is a $\cal G$-martingale under a measure $\mathbb{Q}^h$ where the growth rate of the underlying risky assets  is given by the market rate $h_t$. We can obtain this result also by formally applying the Feynman-Kac theorem on equation \eqref{eq:fundingwithnettingset} as in \cite{Perini2012}, and then deriving the BSDE formulation as above.

Equation \eqref{eq:bsde} is highly non-linear since the variation and initial margins and the rates $c_t$, $f_t$, $f^{N^C}_t$, $f^{N^I}_t$ and $h_t$ may depend on the derivative price $V$ itself and on its partial derivatives w.r.t. the underlying risky assets. Consider for instance the possibility of asymmetric funding rates (borrowing vs lending) according to the sign of the cash needed to implement the hedging strategy, or a variation margin equal to a fraction of the derivative price. Some explicit examples can be found in \cite{Perini2011,Perini2012} and \cite{BrigoPallavicini2013}. Furthermore, a possible dependency of $h_t$ on the derivative price leads to a forward-backward system of stochastic differential equations (FBSDE) as discussed for instance in \cite{Pardoux1999}. A complete study of the above equation is beside the scope of the present paper, but we can refer the reader to \cite{KPQ1997} for a review on the BSDE formulation of pricing equations.

In the following sections we numerically solve equation \eqref{eq:bsde} in a simpler setting when dealing with interest-rate derivatives cleared by CCPs.

It is useful for the discussion in the next sections to write equation \eqref{eq:bsde} again in terms of an expectation, now under the $\mathbb{Q}^h$ measure. We obtain by formally integrating the BSDE formulation
\begin{eqnarray}
\label{eq:fundingwithnettingsetunderQh}
V_t
& = & \int_t^T \,\ExGT{t}{h}{ \ind{u<\tau} D(t,u;f) ( \Pi(u,u+du) + \ind{\tau\in du} \theta_u ) } \\\nonumber
& + & \int_t^T du \,\ExGT{t}{h}{ \ind{u<\tau} D(t,u;f) ( ( f_u - c_u ) M_u + ( f^{N^C}_u - c_u ) N^C_u + ( f^{N^I}_u - c_u ) N^I_u) }
\end{eqnarray}%
where we define the expression
\[
D(t,T;x) := \exp\left\{-\int_t^T du\, x_u \right\} \,.
\]%

In the next section we continue the discussion by inspecting the closeout cash flow $\theta_\tau$ occurring in the terminal condition in case of early default.

\subsection{Close-Out Netting Rules and Gap Risk}
\label{sec:cvadva}

In case of early default of the client ``C", the clearing member ``I" takes  responsibility for the position as in a bilateral trade. The default procedures may take $\delta$ days to be completed (margin period of risk), and in such time-frame the mark-to-market value of the derivative may change considerably, leading to an exposure showing a large mismatch with the value of collateral assets.

Since we have a delay between the first default time $\tau$ and the end of the default procedure at $\tau+\delta$, we should consider the possibility that also the surviving party may default in this lapse of time. In such case we assume that the first default procedure stays in place with the surviving party substituted by his bankruptcy trustee, so that the cash flows originating from such party may be reduced by a recovery rate.

The cash flows occurring in the bilateral case are described in \cite{BrigoCapponiPallaviciniPapatheodorou}. Here, we extend that analysis by considering both initial and variation margins, and by introducing explicitly the margin period of risk. 

\subsection{Margin Period of Risk}
\label{sec:mpr}

When the counterparty (or the client in the CCP case) "C" is the first name to default, say at time $\tau$, the investor (the clearing member) "I" evaluates the exposure $\varepsilon$, also known as close-out amount. This valuation is defined by searching for a replacement deal on the market, as described by the relevant ISDA documentation \cite{ISDA2010}. This documentation specifies that the surviving party may take into account the costs of terminating, liquidating or re-establishing any hedge or related trading position and, furthermore, can consider the cost of funding. A more detailed discussion can be found in \cite{BrigoCapponiPallaviciniPapatheodorou}.

In particular, we can describe the default procedure by assuming that at default time $\tau$ the surviving party enters a deal with a cash flow $\vartheta$, at maturity time $\tau+\delta$, depending on the close-out netting rules. These rules will depend also on the close-out amount $\varepsilon_{\tau+\delta}$ and on the value of the variation and initial margin accounts just before the default event, namely on $M_{\tau^-}$, $N^C_{\tau^-}$, and $N^I_{\tau^-}$. The deal is funded by the surviving party, so that we can use \eqref{eq:fundingwithnettingsetunderQh} to calculate the price $\theta_{\tau}$ and we get 
\[
\theta_{\tau} := \ExGT{\tau}{h}{ D(\tau,\tau+\delta;f^S) \vartheta_{\tau+\delta}\left(\varepsilon_{\tau+\delta},M_{\tau^-},N^C_{\tau^-},N^I_{\tau^-}\right) }
\]%
where $f^S_t$ is the funding rate of the surviving party, and we have dropped the default time indicators since we are assuming  that, in case of further default of the surviving party, this is substituted by his bankruptcy trustee in the default procedure. Notice that in the case of a CCP cleared contract we have $N^I_{\tau^-}=0$, whatever the default time, since the clearing member does not post the initial margin. 

The above result relies on knowledge of $f^S_t$ which, in turn, implies that the price depends on the funding rates of both parties. Yet, the margin period of risk is a time interval of a few days (five for a CCP, and ten for a CSA based bilateral contract), so that we can safely approximate the discount factors by assuming that the payment happens at $\tau$ without modelling the funding rates of both parties. Furthermore, we could compare this approximation with the more relevant uncertainties associated with the recovery rates and the close-out values, concluding that the effects of this approximation are second order effects. Thus, in the following we simply write
\begin{equation}
\label{eq:theta} 
\theta_{\tau} := \ExGT{\tau}{h}{ \vartheta_{\tau+\delta}\left(\varepsilon_{\tau+\delta},M_{\tau^-},N^C_{\tau^-},N^I_{\tau^-}\right) } \,
\end{equation}%

In the above equation we need to evaluate the margin accounts just before the default time. Since the margining procedure is on a discrete time grid, we need to define the account values between two margining dates. By following \cite{Perini2011} we consider that each account accrues continuously at collateral rate $c_t$.

\subsection{The Default Procedure}

We can now investigate the close-out netting rules to calculate $\vartheta_{\tau+\delta}$ in terms of the close-out amount and collateralization accounts. We have different situations according to the sign of the close-out amount and the collateral accounts\footnote{We recall that we consider all the cash flows from the point of view of the investor ``I'' (or the clearing member in the CCP case), and we define in our notation $X^+ := \max(X,0), \ \ X^- := \min(X,0)$.}. We list them below in scenarios where the counterparty ``C'' (the client in the CCP case) defaults first, i.e. when $\ind{\tau_C<\tau_I}$ holds, the other cases being analogous. Hence, the four cases below represent all possible four combinations of signs of exposure and variation margins when
\[ \tau=\tau_C<\tau_I, \ \ 
N^C_{\tau^-}\geq 0, \ \ 
N^I_{\tau^-}\leq 0.
   \]

\subsubsection{Positive Exposure and Positive Variation Margin}

We consider a bilateral contract when the investor measures a positive close-out amount on counterparty default, and some collateral posted as variation margin by the counterparty is available, so that
\[
\varepsilon_{\tau+\delta}\geq 0, \ \ 
M_{\tau^-}\geq 0.
\]%

In this case the investor's exposure is reduced by netting with both the initial and the variation margin, and the remaining collateral is returned to the counterparty. In detail, we can differentiate three different situations:
\begin{enumerate}
\item The collateral is not enough to cover the exposure, so that the investor suffers a loss, then he gets back his initial margin:
\[
\ind{\tau_C<\tau_I} \ind{\varepsilon_{\tau+\delta}\geq 0} \ind{M_{\tau^-}\geq 0} \ind{\varepsilon_{\tau+\delta}-M_{\tau^-}\geq N^C_{\tau^-}} ( \rec_C (\varepsilon_{\tau+\delta} - M_{\tau^-} - N^C_{\tau^-}) - N^I_{\tau^-}) \,.
\]%
\item The collateral is enough to cover the exposure if both the variation margin and counterparty's initial margin are used, so that the investor does not suffer any loss, then he gets back his initial margin:
\[
\ind{\tau_C<\tau_I} \ind{\varepsilon_{\tau+\delta}\geq 0} \ind{M_{\tau^-}\geq 0} \ind{\varepsilon_{\tau+\delta}-M_{\tau^-}<N^C_{\tau^-}} (\varepsilon_{\tau+\delta} - M_{\tau^-} - N^C_{\tau^-} - N^I_{\tau^-}) \,.
\]%
\item The collateral is enough to cover the exposure by using only the variation margin, so that the investor does not suffer any loss. The investor gets back his initial margin unless he defaults before the end of the margin period of risk. In such case the investor's initial margin can be used to reduce losses. Thus, if the investor defaults after the margin period of risk, we have
\[
\ind{\tau_C+\delta<\tau_I} \ind{\varepsilon_{\tau+\delta}\geq 0} \ind{M_{\tau^-}\geq 0} \ind{\varepsilon_{\tau+\delta}-M_{\tau^-}<0} ( \varepsilon_{\tau+\delta} - M_{\tau^-} - N^C_{\tau^-} - N^I_{\tau^-}) \,,
\]%
while, if the investor defaults within the margin period of risk, we have
\begin{eqnarray*}
&& \ind{\tau_C<\tau_I<\tau_C+\delta} \ind{\varepsilon_{\tau+\delta}\geq 0} \ind{M_{\tau^-}\geq 0} \ind{\varepsilon_{\tau+\delta}-M_{\tau^-}<0} (\varepsilon_{\tau+\delta} - M_{\tau^-} - N^I_{\tau^-})^+ \\
& + & \ind{\tau_C<\tau_I<\tau_C+\delta} \ind{\varepsilon_{\tau+\delta}\geq 0} \ind{M_{\tau^-}\geq 0} \ind{\varepsilon_{\tau+\delta}-M_{\tau^-}<0} \rec'_I (\varepsilon_{\tau+\delta} - M_{\tau^-} - N^I_{\tau^-})^- \\
& - & \ind{\tau_C<\tau_I<\tau_C+\delta} \ind{\varepsilon_{\tau+\delta}\geq 0} \ind{M_{\tau^-}\geq 0} \ind{\varepsilon_{\tau+\delta}-M_{\tau^-}<0} N^C_{\tau^-} \,.
\end{eqnarray*}%
\end{enumerate}

\subsubsection{Positive Exposure and Negative Variation Margin}

We continue with a bilateral contract when the investor measures a positive close-out amount on counterparty default, and some collateral posted as variation margin by the investor is available, so that
\[
\varepsilon_{\tau+\delta}>0, \ \ 
M_{\tau^-}<0.
\]%

In this case the investor's exposure is reduced by netting with the initial margin, while the variation margin is returned to the counterparty if it is not re-hypothecated, otherwise only a recovery fraction of it is returned. In detail, we can distinguish two different situations:
\begin{enumerate}
\item The counterparty's initial margin is not enough to cover the exposure, so that the investor suffers a loss, then he gets back his initial margin:
\[
\ind{\tau_C<\tau_I} \ind{\varepsilon_{\tau+\delta}\geq 0} \ind{M_{\tau^-}<0} \ind{\varepsilon_{\tau+\delta}\geq N^C_{\tau^-}} ( \rec_C (\varepsilon_{\tau+\delta}  - N^C_{\tau^-}) - \rec'_C M_{\tau^-} - N^I_{\tau^-}) \,.
\]%
\item The counterparty's initial margin is enough to cover the exposure, so that the investor does not suffer any loss, unless the variation margin is re-hypothecated, then he gets back his initial margin:
\[
\ind{\tau_C<\tau_I} \ind{\varepsilon_{\tau+\delta}\geq 0} \ind{M_{\tau^-}<0} \ind{\varepsilon_{\tau+\delta}<N^C_{\tau^-}} ( (\varepsilon_{\tau+\delta} - M_{\tau^-} - N^C_{\tau^-})^- + \rec'_C (\varepsilon_{\tau+\delta} - M_{\tau^-} - N^C_{\tau^-})^+ - N^I_{\tau^-}) \,.
\]%
Notice that in this case we have not to further differentiate according to the default of the investor, since he holds only the counterparty's initial margin, which is segregated, and not the variation margin, so that there are not losses arising from the investor's default within the margin period of risk.
\end{enumerate}

\subsubsection{Negative Exposure and Positive Variation Margin}

We consider a bilateral contract when the investor measures a negative close-out amount on counterparty default, and some collateral posted as variation margin by the counterparty is available, so that
\[
\varepsilon_{\tau+\delta}<0, \ \ 
M_{\tau^-}\geq 0.
\]%

In this case the exposure is paid to the counterparty, and the counterparty gets back its collateral in full, unless the investor defaults within the margin period of risk. Thus, if the investor defaults after the margin period of risk, we have
\[
\ind{\tau_C+\delta<\tau_I} \ind{\varepsilon_{\tau+\delta}<0} \ind{M_{\tau^-}\geq 0} ( \varepsilon_{\tau+\delta} - M_{\tau^-} - N^C_{\tau^-} - N^I_{\tau^-}) \,,
\]%
while, if the investor defaults within the margin period of risk, we have
\begin{eqnarray*}
&& \ind{\tau_C<\tau_I<\tau_C+\delta} \ind{\varepsilon_{\tau+\delta}<0} \ind{M_{\tau^-}\geq 0} \rec_I (\varepsilon_{\tau+\delta} - N^I_{\tau^-})^- \\
& + & \ind{\tau_C<\tau_I<\tau_C+\delta} \ind{\varepsilon_{\tau+\delta}<0} \ind{M_{\tau^-}\geq 0} \rec'_I ((\varepsilon_{\tau+\delta} - N^I_{\tau^-})^+ - M_{\tau^-})^- \\
& + & \ind{\tau_C<\tau_I<\tau_C+\delta} \ind{\varepsilon_{\tau+\delta}<0} \ind{M_{\tau^-}\geq 0} ((\varepsilon_{\tau+\delta} - N^I_{\tau^-})^+ - M_{\tau^-})^+ \\
& - & \ind{\tau_C<\tau_I<\tau_C+\delta} \ind{\varepsilon_{\tau+\delta}<0} \ind{M_{\tau^-}\geq 0} N^C_{\tau^-} \,.
\end{eqnarray*}%
The above formula is simply the contributions occurring in the case discussed in the previous case when the investor is the first defaulting name.

\subsubsection{Negative Exposure and Negative Variation Margin}

We conclude by discussing a bilateral contract when the investor measures a negative close-out amount on counterparty default, and some collateral posted as variation margin by the investor is available, so that
\[
\varepsilon_{\tau+\delta}<0, \ \ 
M_{\tau^-}<0.
\]%

In this case the initial margin is used to reduce the possible losses coming from a re-hypothecated variation margin. In detail, by following the same scheme of the first case, we can differentiate three different situations:
\begin{enumerate}
\item The counterparty's initial margin is not enough to cover the losses due to the re-hypothecation of the variation margin, so that the investor  suffers a loss, then he gets back his initial margin:
\[
\ind{\tau_C<\tau_I} \ind{\varepsilon_{\tau+\delta}<0} \ind{M_{\tau^-}<0} \ind{\varepsilon_{\tau+\delta}-M_{\tau^-}\geq N^C_{\tau^-}} ( \rec'_C (\varepsilon_{\tau+\delta} - M_{\tau^-} - N^C_{\tau^-}) - N^I_{\tau^-}) \,.
\]%
\item The counterparty's initial margin is enough to cover the losses due to the re-hypothecation of the variation margin, so that the investor does not suffer any loss, then he gets back his initial margin:
\[
\ind{\tau_C<\tau_I} \ind{\varepsilon_{\tau+\delta}<0} \ind{M_{\tau^-}<0} \ind{\varepsilon_{\tau+\delta}-M_{\tau^-}<N^C_{\tau^-}} (\varepsilon_{\tau+\delta} - M_{\tau^-} - N^C_{\tau^-} - N^I_{\tau^-}) \,.
\]%
\item The variation margin is not returned to the investor since he has to pay a greater exposure. The investor gets back his initial margin unless he defaults before the end of the margin period of risk. In such case the investor's initial margin can be used to reduce losses. Thus, if the investor defaults after the margin period of risk, we have
\[
\ind{\tau_C+\delta<\tau_I} \ind{\varepsilon_{\tau+\delta}<0} \ind{M_{\tau^-}<0} \ind{\varepsilon_{\tau+\delta}-M_{\tau^-}<0} ( \varepsilon_{\tau+\delta} - M_{\tau^-} - N^C_{\tau^-} - N^I_{\tau^-}) \,,
\]%
while, if the investor defaults within the margin period of risk, we have
\begin{eqnarray*}
&& \ind{\tau_C<\tau_I<\tau_C+\delta} \ind{\varepsilon_{\tau+\delta}<0} \ind{M_{\tau^-}<0} \ind{\varepsilon_{\tau+\delta}-M_{\tau^-}<0} (\varepsilon_{\tau+\delta} - M_{\tau^-} - N^I_{\tau^-})^+ \\
& + & \ind{\tau_C<\tau_I<\tau_C+\delta} \ind{\varepsilon_{\tau+\delta}<0} \ind{M_{\tau^-}<0} \ind{\varepsilon_{\tau+\delta}-M_{\tau^-}<0} \rec_I (\varepsilon_{\tau+\delta} - M_{\tau^-} - N^I_{\tau^-})^- \\
& - & \ind{\tau_C<\tau_I<\tau_C+\delta} \ind{\varepsilon_{\tau+\delta}<0} \ind{M_{\tau^-}<0} \ind{\varepsilon_{\tau+\delta}-M_{\tau^-}<0} N^C_{\tau^-} \,.
\end{eqnarray*}%
\end{enumerate}

\subsection{Cash Flows on Default Event}
\label{sec:ondefault}

We have now to sum the above cash flows based on the counterparty defaulting first (the client in the CCP case) with the similar cash flows occurring when the investor (the clearing member) is the first name to default, each cash flow embedding the relevant default indicators. The latter case originates in the bilateral CSA case the same default procedure seen from the opposite side, so that the cash flows can be deduced by symmetry arguments as in \cite{BrigoCapponiPallaviciniPapatheodorou}, while in the CCP case a different default procedure takes place and it is managed by the clearing house. However, from the point of view of the cash flows we can do the calculations as in the bilateral case by exchanging the clearing member and the client names in the above equations while changing at the same time the sign of all cash flows.

Once the cash flows exchanged on $\tau+\delta$ are known, we can evaluate $\vartheta_{\tau+\delta}$, and in turn $\theta_{\tau}$. If we sum all the cash flows along with the values of the variation and initial margin accounts just before the default event, and we add the sum of all margin accounts just before the default event, we get
\begin{eqnarray}
\label{eq:theta_explicit}
\theta_{\tau}^{\rm bilateral}
&=& \ExGT{\tau}{h}{ \varepsilon_{\tau+\delta}} \\\nonumber
&-& \,\ExGT{\tau}{h}{ \ind{\tau_C<\tau_I+\delta} \lgd_C ((\varepsilon_{\tau+\delta}-N^C_{\tau^-})^+ - M_{\tau^-}^+)^+} \\\nonumber
&-& \,\ExGT{\tau}{h}{ \ind{\tau_C<\tau_I+\delta} \lgd'_C ((\varepsilon_{\tau+\delta}-N^C_{\tau^-})^- - M_{\tau^-}^-)^+} \\\nonumber
&-& \,\ExGT{\tau}{h}{ \ind{\tau_I<\tau_C+\delta} \lgd_I ((\varepsilon_{\tau+\delta}-N^I_{\tau^-})^- - M_{\tau^-}^-)^-} \\\nonumber
&-& \,\ExGT{\tau}{h}{ \ind{\tau_I<\tau_C+\delta} \lgd'_I ((\varepsilon_{\tau+\delta}-N^I_{\tau^-})^+ - M_{\tau^-}^+)^-}
\end{eqnarray}%
where the first term is the replacement price of the deal, as seen at default time, while the second and third terms are the counterparty risk due to client default (also known as counterparty valuation adjustment or CVA), and come with a negative sign (always from the point of view of the clearing member). The fourth and fifth terms represent the counterparty risk due to clearing member default (also known as debit valuation adjustment or DVA) and come with a positive sign (again from the point of view of the clearing member).

We notice that the indicators appearing in equation \eqref{eq:theta_explicit} are not mutually exclusive, but they can be all different from zero when the time interval between the two default times is less than the margin period of risk. Indeed, the CVA terms contribute to the final cash flows even if the first name that defaults is the clearing member, provided that the client defaults before the end of the default procedure. A similar situation arises for the DVA terms too.

Then, if we assume that in case of re-hypothecation of the variation margin the recovery rate for the collateral is the same used for the derivative, we can write
\begin{eqnarray}
\label{eq:theta_rehyp}
\theta^{\rm rehyp}_{\tau}
&:=& \ExGT{\tau}{h}{ \varepsilon_{\tau+\delta}} \\\nonumber
&-& \,\ExGT{\tau}{h}{ \ind{\tau_C<\tau_I+\delta} \lgd_C (\varepsilon_{\tau+\delta} - N^C_{\tau^-} - M_{\tau^-})^+} \\\nonumber
&-& \,\ExGT{\tau}{h}{ \ind{\tau_I<\tau_C+\delta} \lgd_I (\varepsilon_{\tau+\delta} - N^I_{\tau^-} - M_{\tau^-})^-} \,.
\end{eqnarray}%

On the other hand, if re-hypothecation is forbidden, we can set the collateral recovery rate to one, and we obtain
\begin{eqnarray}
\label{eq:theta_norehyp}
\theta^{\rm no-rehyp}_{\tau}
&:=& \ExGT{\tau}{h}{ \varepsilon_{\tau+\delta}} \\\nonumber
&-& \,\ExGT{\tau}{h}{ \ind{\tau_C<\tau_I+\delta} \lgd_C ((\varepsilon_{\tau+\delta}-N^C_{\tau^-})^+ - M_{\tau^-}^+)^+} \\\nonumber
&-& \,\ExGT{\tau}{h}{ \ind{\tau_I<\tau_C+\delta} \lgd_I ((\varepsilon_{\tau+\delta}-N^I_{\tau^-})^- - M_{\tau^-}^-)^-} \,.
\end{eqnarray}%

{
In the case of a CCP cleared contract, we can set $N^I\doteq 0$ into equation \eqref{eq:theta_explicit}, since the initial margin is posted only by the client. Furthermore, if we can assume with certainty that upon clearing member default the position is transferred to a backup clearing member, so that we set the recovery rate for the clearing member to one, then we can completely drop the third term on the right-hand side of equations \eqref{eq:theta_rehyp} and \eqref{eq:theta_norehyp}.
\begin{eqnarray}
\label{eq:theta_ccp}
\theta^{\rm CCP}_{\tau}
&:=& \ExGT{\tau}{h}{ \varepsilon_{\tau+\delta}} \\\nonumber
&-& \,\ExGT{\tau}{h}{ \ind{\tau_C<\tau_I+\delta} \lgd_C (\varepsilon_{\tau+\delta} - N^C_{\tau^-} - M_{\tau^-})^+} \\\nonumber
&-& \,\ExGT{\tau}{h}{ \ind{\tau_I<\tau_C+\delta} \lgd_I (\varepsilon_{\tau+\delta} - M_{\tau^-})^-} \\\nonumber
&\approx& \ExGT{\tau}{h}{ \varepsilon_{\tau+\delta} - \ind{\tau_C<\tau_I+\delta} \lgd_C (\varepsilon_{\tau+\delta} - N^C_{\tau^-} - M_{\tau^-})^+} \,.
\end{eqnarray}%
}

In the following we assume that the variation margin can be re-hypothecated, so that we can substitute the on-default cash flows given by equation \eqref{eq:theta_rehyp} in the pricing equation \eqref{eq:fundingwithnettingsetunderQh} to get
\begin{eqnarray}
\label{eq:funding}
V_t
& = & \int_t^T \,\ExGT{t}{h}{ \ind{u<\tau} D(t,u;f) \Pi(u,u+du) } \\\nonumber
& + & \int_t^T du \,\ExGT{t}{h}{ \ind{u<\tau} D(t,u;f) ( ( f_u - c_u ) M_u + ( f^{N^C}_u - c_u ) N^C_u ) + ( f^{N^I}_u - c_u ) N^I_u ) } \\\nonumber
& + & \int_t^T \,\ExGT{t}{h}{ \ind{\tau\in du} D(t,u;f) \varepsilon_{u+\delta} } \\\nonumber
& - & \int_t^T \,\ExGT{t}{h}{ \ind{\tau\in du} \ind{\tau_C<\tau_I+\delta} D(t,u;f) \lgd_C (\varepsilon_{u+\delta}-N^C_{u^-} - M_{u^-})^+ } \\\nonumber
& - & \int_t^T \,\ExGT{t}{h}{ \ind{\tau\in du} \ind{\tau_I<\tau_C+\delta} D(t,u;f) \lgd_I (\varepsilon_{u+\delta}-N^I_{u^-} - M_{u^-})^- } \,.
\end{eqnarray}%
{ 
We obtain a pricing equation \eqref{eq:funding} which can be used both for bilateral and CCP cleared trades. In the CCP case we assume $N^I_t=0$ and a loss given default for the investor close to zero, as we show in the numerical section.
}

\subsection{Close-Out Amount Evaluation}

According to \cite{ISDA2010}, the surviving party searches for a replacement deal on the market to evaluate the close-out amount, and he can include any costs of terminating, liquidating or re-establishing any hedge or related trading position, along with the cost of funding. Such definition is supported by ISDA documentation, but it is difficult to implement, since we do not know the counterparty of the replacement deal, and, as a consequence, we cannot seize his credit charge into the pricing equations in a precise way. In the literature many recipes are adopted to circumvent this problem. We address the reader to \cite{BrigoCapponiPallaviciniPapatheodorou}, \cite{BurgardKjaer2011b}, or \cite{Durand2013} for technical discussions, and to \cite{BrigoMoriniPallavicini2012} for a review.

Here, we focus on a particular recipe: we assume that the close-out amount is equal to the mark-to-market of the derivative contract considered between two default-free counterparties in case of perfect collateralization. This solution can be considered as a generalization of the risk-free close-out often used in the literature. However, different recipes may be adopted within our framework.

\begin{remark} {\bf (Close-Out Amount and Funding Costs).}
According to the \cite{ISDA2010} documentation the surviving party can include funding costs within the close-out amount evaluation. Thus, we could search for a close-out amount of the form
\[
\varepsilon_{\tau}(f^S) = \ind{\tau=\tau_I<\tau_C} \varepsilon_{\tau_I}(f^I) + \ind{\tau=\tau_C<\tau_I} \varepsilon_{\tau_C}(f^C)
\]%
where we have explicitly written the dependency on funding rates. Notice that the funding rates of the surviving party are used in the estimation of the above amount. Such approach has the drawback of including into the pricing equation the funding costs of both  parties, so that each party has to model the funding costs of the other party, whose Treasury policies may be unknown to the calculating party. A possible solution is considering the CDS/bond basis as a proxy for the unknown funding spreads. Alternatively, we could approximate funding spreads with the funding spreads of the calculating party by exchanging funding and investing rates, since positive cash flows for one party are negative cash flows for the other party and vice versa.
\end{remark}

When dealing with contracts cleared via CCPs we have to consider also the impact of the margin period of risk. Thus, we extend the definition of close-out amount by introducing a delay in the default procedure. In such a case we need to include in the value of the close-out also the cash flows occurring during the delay. Accordingly, we define the close-out amount not as a mark-to-market evaluated at the first default time, but as a mark to market P\&L $\varepsilon_{\tau+\delta}(\tau,T)$ observed at the end of the default procedure $\tau+\delta$ inclusive of delay $\delta$ and starting from the first default time $\tau$. Thus, we write
\begin{equation}
\label{eq:riskfree-closeout}
\varepsilon_{\tau+\delta}(\tau,T) := \int_{\tau}^T \ExGT{\tau+\delta}{h}{ D(\tau,u;c) \Pi(u,u+du) } \,.
\end{equation}%

Furthermore, it is useful to extend the definition of the close-out amount for any time $t$ and $s$ with $t\leq s\leq T$ as given by
\[
\varepsilon_s(t,T) := \int_t^T \ExGT{s}{h}{ D(t,u;c) \Pi(u,u+du) }
\]%
which is equal to the previous definition when evaluated in $t=\tau$ and $s=\tau+\delta$, and we are able to write
\[
\varepsilon_t(t,T) = \int_t^T \ExGT{t}{h}{ D(t,u;f) \left( \Pi(u,u+du) + (f_u - c_u) \varepsilon_u(u,T) du \right) }
\]%
for any funding rate $f_t$.

We can substitute equation \eqref{eq:riskfree-closeout} into equation \eqref{eq:funding}, by taking into account the above results, to obtain
\begin{eqnarray}
\label{eq:fundingexplicit}
V_t
& = & \varepsilon_t(t,T) \\\nonumber
& + & \int_t^T du \,\ExGT{t}{h}{ \ind{u<\tau} D(t,u;f) ( ( f^{N^C}_u - c_u ) N^C_u + ( f^{N^I}_u - c_u ) N^I_u ) } \\\nonumber
& - & \int_t^T \,\ExGT{t}{h}{ \ind{\tau\in du} \ind{\tau_C<\tau_I+\delta} \lgd_C D(t,u;f) \Delta^{C,+}_{u+\delta} } \\\nonumber
& - & \int_t^T \,\ExGT{t}{h}{ \ind{\tau\in du} \ind{\tau_I<\tau_C+\delta} \lgd_I D(t,u;f) \Delta^{I,-}_{u+\delta} }
\end{eqnarray}%
where we define the gap risk measured on counterparty's default $\Delta^C_{\tau+\delta}$ and on investor's default $\Delta^I_{\tau+\delta}$ as
\begin{eqnarray}
\label{eq:gaprisk_whole}
\Delta^C_{\tau+\delta} := \varepsilon_{\tau+\delta}(\tau,T) - N^C_{\tau^-} - M_{\tau^-}
\;,\quad
\Delta^I_{\tau+\delta} := \varepsilon_{\tau+\delta}(\tau,T) - N^I_{\tau^-} - M_{\tau^-} \,.
\end{eqnarray}%

The pricing equation \eqref{eq:fundingexplicit} can be viewed as a generalization of the usual CVA/DVA pricing equation, see for instance \cite{BrigoMoriniPallavicini2012}. Indeed we can define a collateralized CVA/DVA adjustment in presence of funding and hedging costs by subtracting $\varepsilon_t(t,T)$ from both sides. Yet, the above equation is only a formal description of a BSDE problem, and it cannot be solved explicitly, since rates and collateral accounts on the right-hand side may depend on the future prices $V$ of the derivative itself, see Section \ref{sec:hifc} below.

In the following section we focus on gap risk and on how we can reduce it by suitably defining the variation and initial margins.

\subsection{Gap Risk and Margining Procedures}
\label{sec:gaprisk}

Gap risk may be almost instantaneous, building on strong contagion after a default, or may build during the margin period of risk, due to high volatility of the underlying mark to market. The relevance of gap risk depends on the asset class we are considering. For instance, as shown in \cite{BrigoCapponiPallavicini}, CDS prices are heavily affected by instantaneous gap risk, since the mark-to-market of a CDS jumps when one of the counterparties defaults. This jump occurs because of the dependence between the default time of the CDS reference name and the default time of the defaulted counterparty. The initial margins requested by a CCP protect from such risk, while the variation margins protect from the loss of exposure as measured at the last margin call.

We can highlight the different contributions hinted above by rewriting the gap risk definition in the following form (we omit the superscripts on gap risks and initial margins to lighten notations).

\begin{equation}
\label{eq:gaprisk}
\Delta_{\tau+\delta} = \Delta^{\rm\scriptscriptstyle Mismatch}_{\tau^-} + \Delta^{\rm\scriptscriptstyle Contagion}_{\tau}  + \Delta^{\rm\scriptscriptstyle MtM}_{\tau+\delta}
\end{equation}%
where we define three components of the gap risk according to three different fixing times: (i) just before the default event, (ii) the default event, and (iii) the end of the default procedure. We list them below in detail.
\begin{enumerate}
\item The component of the gap risk due to a mismatch between the variation margin account and the value of the close-out amount, if it was calculated just before the default event, is given by
\begin{equation}
\label{eq:gaprisk_mismatch}
\Delta^{\rm\scriptscriptstyle Mismatch}_{\tau^-} := \varepsilon_{\tau^-}(\tau,T) - M_{\tau^-} \,.
\end{equation}%
\item The component of the gap risk due to an instantaneous contagion effect at default time is given by
\begin{equation}
\label{eq:gaprisk_contagion}
\Delta^{\rm\scriptscriptstyle Contagion}_{\tau} := \varepsilon_{\tau}(\tau,T) - \varepsilon_{\tau^-}(\tau,T) - N^{\rm\scriptscriptstyle Contagion}_{\tau^-} \,.
\end{equation}%
\item The component of the gap risk due to a movement in the mark-to-market of the close-out amount between the default event and the end of the default procedure is given by
\begin{equation}
\label{eq:gaprisk_mtm}
\Delta^{\rm\scriptscriptstyle MtM}_{\tau+\delta} := \varepsilon_{\tau+\delta}(\tau,T) - \varepsilon_{\tau}(\tau,T) - N^{\rm\scriptscriptstyle MtM}_{\tau^-} \,.
\end{equation}%
\end{enumerate}
where we can split the initial margin contribution into two parts:
\[
N_{\tau^-} := N^{\rm\scriptscriptstyle Contagion}_{\tau^-} + N^{\rm\scriptscriptstyle MtM}_{\tau^-} \,.
\]%
This partition is implicit but we will make it explicit in the following examples on interest rate derivatives. 
We observe that both the variation and the initial margin are fixed just before the default event, namely at $\tau^-$, as the mismatch term of gap risk is fixed, while the contagion and the mark-to-market terms of gap risk are fixed at later times, respectively at $\tau$ and at $\tau+\delta$. Thus, we are able only to drop the mismatch term of gap risk by a suitable definition of the variation margin, while the other two terms can be reduced but not eliminated.

In the following section we will discuss how to define the variation and the initial margins so as to reduce the gap-risk contributions in the case of interest rate derivatives cleared by a CCP or traded in a bilateral contract under CSA.

\section{Pricing Interest-Rate Derivatives under Bilateral CSA or CCP Clearing}\label{sec:IRD}

In the previous section we derived the pricing equation inclusive of counterparty credit risk and funding costs for a generic derivative in presence of variation and initial margins by taking into account the margin period of risk. Here, we wish to specialize the problem to the case of interest-rate derivatives. As a first step we introduce a collateralization procedure suitable for interest-rate derivatives, then we discuss how to model the rates appearing in the pricing equations, so as to write a pricing equation we can numerically solve.

\subsection{Variation and Initial Margin Estimates}

The variation margin can be defined to include all the derivative cash flows with the exception of those depending on the funding costs of the counterparties. In the case of interest-rate derivatives cleared by a CCP, we usually find contractual rules explaining how to calculate the variation margin by discounting the derivative coupons at an official rate issued by the CCP at the end of each trading day. Here, we assume that such rate is the collateral rate $c_t$, so that we can define
\begin{equation}
\label{eq:variation_margin}
M_t \doteq \varepsilon_t(t,T)
\end{equation}%
and drop the mismatch gap-risk term, namely we get
\[
\Delta^{{\rm\scriptscriptstyle Mismatch},C}_{\tau^-} = \Delta^{{\rm\scriptscriptstyle Mismatch},I}_{\tau^-} = 0 \,.
\]%

\begin{remark}{\bf (Partial Collateralization in Bilateral Trades).}\label{rem:vmfrac}
In bilateral trades when collateralization is not full, it can be useful to define the variation margin as in \cite{BrigoPallavicini2013} where it is set equal to a fraction $\alpha_t$ of the close-out, namely we define
\begin{equation}
\label{eq:variation_margin_frac}
M^{\rm partial}_t \doteq \alpha_t \varepsilon_t(t,T)
\end{equation}%
We employ this definition in the numerical examples of Section \ref{sec:numerics} to show the transition between non-collateralized and (fully-)collateralized deals.
\end{remark}

On the other hand, initial margins can be estimated by the CCP according to the prevailing market conditions and the expected maturity of the portfolio. The gap risk arising from the mark-to-market term is usually analyzed in terms of historical Value-at-Risk (VaR) or Expected Shortfall (ES) estimates. Furthermore, CCPs can apply multipliers to initial margins at their discretion. For instance, they can call additional margins to cover potential losses arising from pricing impacts in executing large trades in case of a member default, or when a member suffers a downgrade.

Here, we are interested in pricing interest-rate derivatives. According to the analysis of \cite{BrigoCapponiPallaviciniPapatheodorou} we can safely disregard the contagion component of gap risk for interest-rate derivatives. Technically, this is to say that an interest rate exposure  does not jump by contagion at default of a counterparty. Thus, we can assume that
\[
\Delta^{{\rm\scriptscriptstyle Contagion},C}_{\tau} \doteq \Delta^{{\rm\scriptscriptstyle Contagion},I}_{\tau} \doteq 0
\]%
and
\begin{equation}
\label{eq:initial_margin_ir}
N^{{\rm\scriptscriptstyle Contagion},C}_{\tau^-} \doteq N^{{\rm\scriptscriptstyle Contagion},I}_{\tau^-} \doteq 0
\,.
\end{equation}%
so that
\[
N^C_{\tau^-} = N^{{\rm\scriptscriptstyle MtM},C}_{\tau^-}
\;,\quad
N^I_{\tau^-} = N^{{\rm\scriptscriptstyle MtM},I}_{\tau^-}
\,.
\]%

Then we can calculate the initial margin posted to protect from mark-to-market movements as the protection against the worst movement of the contract due to market risk within $\delta$ days at a confidence level $q$ according to some risk measure, typically Value at Risk or Expected Shortfall. Again we use the rate issued by the CCP to calculate mark-to-market and P\&L. For instance, if we assume as risk measure the Value at Risk (VaR) of the price variation of the contract we obtain that the initial margins $N^C_t$ and $N^I_t$ can be defined as given by
\[
N^{{\rm\scriptscriptstyle MtM},C}_t \doteq \inf\left\{ x\geq 0 : \PxC{\varepsilon_{t+\delta}(t,T) - \varepsilon_t(t,T) < x}{{\cal F}_t} > q \right\}
\]%
and only for bilateral contracts under CSA
\[
N^{{\rm\scriptscriptstyle MtM},I}_t \doteq \sup\left\{ x\leq 0 : \PxC{\varepsilon_{t+\delta}(t,T) - \varepsilon_t(t,T) > x}{{\cal F}_t} > q \right\}
\]%
where the $\mathbb{P}$ is the physical probability measure, and we have modified the VaR definition to ensure that the initial margin is posted only by the one counterparty at time. In a CCP cleared contract we consider only the initial margin posted by the client, while in a bilateral transaction the proposal of ISDA \cite{ISDA2013} states that each counterparty has his own initial margin to post to the other one without any netting assumption. An analogous definition can be formulated with ES replacing VaR in the above equation.

In the following we discuss numerical examples, and we need to evaluate the initial margin account within the simulation of the pricing equation. Thus, we face the problem of evaluating a quantile under the physical measure $\mathbb{P}$ while simulating under the pricing measure $\mathbb{Q}^h$. Our proposal is to approximate the calculation by replacing the above recipe for the initial margin calculated in the physical measure by the corresponding one in the pricing measure, namely we define
\begin{equation}
\label{eq:initial_margin_C}
N^{{\rm\scriptscriptstyle MtM},C}_t \doteq \inf\left\{ x\geq 0 : \QxCT{h}{\varepsilon_{t+\delta}(t,T) - \varepsilon_t(t,T) < x}{{\cal F}_t} > q \right\}
\end{equation}%
{  and only for bilateral contracts under CSA
\begin{equation}
\label{eq:initial_margin_I}
N^{{\rm\scriptscriptstyle MtM},I}_t \doteq \sup\left\{ x\leq 0 : \QxCT{h}{\varepsilon_{t+\delta}(t,T) - \varepsilon_t(t,T) > x}{{\cal F}_t} > q \right\}
\end{equation}%
} where we can take the expectation conditional on the default-free market filtration, since for interest-rate payoffs we do not have a dependency of the payout on default times of reference names. 

We can further approximate the initial margin in order to speed up the pricing simulation. Indeed, we can implement a moment matching technique to approximate the conditional probability, appearing in the VaR calculation, with a normal distribution
\begin{equation}
\label{eq:initial_margin_approx_C}
N^{{\rm\scriptscriptstyle MtM},C}_t \approx \Phi^{-1}_{0,\nu_t}(q)
\end{equation}%
where $\Phi_{\mu,\nu}$ is the cumulative distribution function of a normal distribution with mean $\mu$ and standard deviation $\nu$, and where the mean and variance are calculated under the  pricing measure w.r.t. the default-free market filtration, and are given by
\[
\mu_t := 0
\;,\quad
\nu^2_t := \VxFT{t}{h}{\varepsilon_{t+\delta}(t,T)} \,.
\]%
Only for bilateral contracts under CSA we can accordingly define the initial margin posted by the investor as given by
\begin{equation}
\label{eq:initial_margin_approx_I}
N^{{\rm\scriptscriptstyle MtM},I}_t \approx - N^{{\rm\scriptscriptstyle MtM},C}_t \,.
\end{equation}%
being zero in the CCP case.

\subsection{Collateralized Hedging Instruments and Funding Costs}\label{sec:hifc}

The money market usually quotes the prices of collateralized instruments, such as standardized interest-rate swaps (IRS) and overnight indexed swaps (OIS). In particular, a daily collateralization procedure is assumed, so that CSA contracts require that the collateral account be remunerated at the overnight rate ($e_t$). The same happens for quoted instruments sensitive to credit spreads of defaultable names, such as the credit default swaps (CDS). Such instruments are usually used as hedging instruments\footnote{A name cannot sell protection on himself by means of his own CDS, so that he must implement a different hedging strategy in the practice. See \cite{BrigoMoriniPallavicini2012} for a discussion with examples.}.

When we hedge a derivative contract by means of collateralized hedging instruments we must consider also the cash flows needed to implement their collateral procedure, as we are already doing for the collateral procedure of the derivative itself. In our framework this can be accomplished by setting the borrowing and lending rates, used to trade the hedging instruments, equal to their collateralization rate, as shown in \cite{BrigoPallavicini2013}. Thus, we can write
\begin{equation}
\label{eq:htilde}
h_t \doteq e_t \,.
\end{equation}%

Then, we can discuss funding rates, namely the rates used to borrow and lend cash. We assume that in general they can be different, and we select the borrowing or the lending rate according to the sign of the netting set that the Treasury assigns to the derivative contract. Since we are using collateralized hedging instruments, we consider as a single funding netting set the set defined by the need for cash of both the derivative position and its hedge, so that we can define the funding rate as given by 
\[
f_t \doteq f^+_t \ind{F_t+C^H_t>0} + f^-_t \ind{F_t+C^H_t<0} \,,
\]%
where $C^H_t$ is the collateral account of the hedging instruments and $F$ is the cash component of the hedge portfolio, which is defined in equation \eqref{eq:replica}. Here, we use the symbol $f^+_t$ to represent the funding borrowing rate, while $f^-_t$ is the lending rate.

For example, as we have seen above, in the Black Scholes model $H_t$ is the value of the delta position in the stock, whereas $F_t$ is the value of the remaining cash component of the replication. Furthermore, in the case of interest-rate derivatives, where we can assume that hedging instruments are collateralized in a way that perfectly matches the mark-to-market over the discrete time margining dates, we have $C^H_t = H_t$, leading to
\begin{equation}
\label{eq:ftilde}
f_t = f^+_t \ind{V_t>M_t} + f^-_t \ind{V_t<M_t} \, .
\end{equation}%

\begin{remark}\label{rem:nonlin} {\bf Fundamental nonlinearity of the pricing operator in funding inclusive valuation.}
{This last equation above is quite important. It shows that the rates $f$ appearing in our pricing problem depend on the future size of the claim value $V$ compared to the collateral $M$, and in turn we know the price value $V$ will depend on future values of the funding cash flows $\varphi$ depending on those same rates $f$. This introduces a fundamental nonlinearity at the level of pricing operators or expectations in the valuation problem, a nonlinearity that was not there at the pure credit valuation adjustment (CVA) level, since CVA induces nonlinearity in the payout but not on the pricing operators or expectations.}
\end{remark}
 
We conclude this section by considering the collateral rates usually adopted by CCPs or by bilateral contracts under CSA (in particular when a SCSA is adopted). In these contracts the collateral rate is usually equal to the overnight rate for both counterparties. Thus, in the following we can write
\begin{equation}
\label{eq:ctilde}
c_t \doteq e_t \,.
\end{equation}

{ 
It is also useful to express the funding and investing rates as a liquidity basis over the overnight rate $e_t$, so that we have
\begin{equation}
\label{eq:fasell}
f^+_t \doteq e_t + \ell^+_t
\;,\quad
f^-_t \doteq e_t + \ell^-_t \,,
\end{equation}
while for the funding rates used for the initial margin accounts we have
\begin{equation}
\label{eq:ftildeN}
f^{N^C}_t \doteq e_t + \ell^{N^C}_t
\;,\quad
f^{N^I}_t \doteq e_t + \ell^{N^I}_t \,.
\end{equation}%
The liquidity bases $\ell^+_t$, $\ell^-_t$, $\ell^{N^C}$ and $\ell^{N^I}$ can be obtained from the market from the bond/CDS basis or modelled according to a stylized version of the Treasury liquidity policy as described in \cite{BrigoPallavicini2013}. In the latter case the liquidity bases may be related to the default intensity and/or to credit worthiness of the collateral portfolio used by the Treasury for funding.

In this paper we assume that the liquidity bases are related to the credit spreads of the calculating party, namely to the pre-default intensities of the two names $\lambda^I_t$ and $\lambda^C_t$, discussed later in equation \eqref{eq:lambda}, via some multipliers $\beta^\pm\in[0,1]$. The multipliers depend on the Treasury policy, and we assume them to be deterministic constants in this initial analysis.

We present two different cases which we analyze in Section \ref{sec:numerics} in detail, according to the identity of the calculating party. Here a complete notation would require us to add a further index to bases and rates to specify which entity ("I" or "C") is doing the calculation. To avoid cumbersome notation we omit this further index. 
\begin{itemize}
\item Prices calculated by the investor.
\begin{equation}
\label{eq:liquidity_basis_I}
\ell^+_t \doteq \beta^+ \lambda^I_t
\;,\quad
\ell^-_t \doteq \beta^- \lambda^I_t
\;,\quad
\ell^{N^C}_t \doteq 0
\;,\quad
\ell^{N^I}_t \doteq \ell^+_t \,.
\end{equation}%
\item Prices calculated by the counterparty.
\begin{equation}
\label{eq:liquidity_basis_C}
\ell^+_t \doteq \beta^+ \lambda^C_t
\;,\quad
\ell^-_t \doteq \beta^- \lambda^C_t
\;,\quad
\ell^{N^C}_t \doteq \ell^-_t
\;,\quad
\ell^{N^I}_t \doteq 0 \,.
\end{equation}%
\end{itemize}
}

\subsection{Dealing with Default Dependencies and Delays}
\label{sec:delay}

We plug into the pricing equation \eqref{eq:fundingexplicit} the estimates of the variation and initial margin, given by equations \eqref{eq:variation_margin} ad \eqref{eq:initial_margin_ir}, along with the modelling choices on market rates of equations \eqref{eq:ctilde}, \eqref{eq:htilde}, \eqref{eq:ftilde}. We obtain
\begin{eqnarray*}
V_t
& = & \varepsilon_t(t,T) \\
& + & \int_t^T du \,\ExGT{t}{e}{ \ind{u<\tau} D(t,u;f) ( \ell^{N^C}_u N^C_u + \ell^{N^I}_u N^I_u ) } \\
& - & \int_t^T \,\ExGT{t}{e}{ \ind{\tau\in du} \ind{\tau_C<\tau_I+\delta} \lgd_C D(t,u;f) (\Delta^C_{u+\delta})^+ } \\
& - & \int_t^T \,\ExGT{t}{e}{ \ind{\tau\in du} \ind{\tau_I<\tau_C+\delta} \lgd_I D(t,u;f) (\Delta^I_{u+\delta})^- } \,.
\end{eqnarray*}%

When the close-out amount is ${\cal F}$-adapted, as in the case of interest-rate derivatives, we can simplify the pricing equation given by \eqref{eq:fundingexplicit} by switching to the default-free market filtration. By following the filtration switching formula in \cite{BrigoMercurio2006}, or the more thorough analysis in  \cite{BCJR2008}, we introduce for any ${\cal G}_t$-adapted process $X_t$ a unique ${\cal F}_t$-adapted process ${\widetilde X}_t$, defined such that
\[
\ind{\tau>t} X_t = \ind{\tau>t} {\widetilde X}_t .
\]%
Thus, we can write the pre-default price process as given by
\begin{eqnarray*}
\ind{\tau>t} {\widetilde V}_t
& = & \ind{\tau>t} \varepsilon_t(t,T) \\
& + & \int_t^T du \,\ExGT{t}{e}{ \ind{u<\tau} D(t,u;f) ( \ell^{N^C}_u N^C_u + \ell^{N^I}_u N^I_u ) } \\
& - & \int_t^T \,\ExGT{t}{e}{ \ind{\tau\in du} \ind{\tau_C<\tau_I+\delta} \lgd_C D(t,u;f) (\Delta^C_{u+\delta})^+ } \\
& - & \int_t^T \,\ExGT{t}{e}{ \ind{\tau\in du} \ind{\tau_I<\tau_C+\delta} \lgd_I D(t,u;f) (\Delta^I_{u+\delta})^- }
\end{eqnarray*}%
where we recall that in our case also the margin accounts and the gap risks are ${\cal F}$-adapted processes.

Our intention is to study the impact of the liquidity bases on interest-rate derivative prices cleared by a CCP, and to extend the analysis of \cite{BrigoCapponiPallaviciniPapatheodorou} on the dependency of derivative prices w.r.t. changes in the correlation between market and credit risks (wrong-way risk). In this first work on CCPs we develop such analysis without considering a dependence between the default times if not through their spreads, or more precisely by assuming that the default times are ${\cal F}$-conditionally independent. 

Thus, we can write for any time $t$ and $u$ such that $t\leq u$
\[
\ExGT{t}{e}{ \ind{\tau\in du} \ind{\tau_C<\tau_I+\delta} \phi_u } = \ind{\tau>t} du\,\ExFT{t}{e}{\lambda^{\delta,C}_u D(t,u;\lambda) \phi_u }
\]%
and
\[
\ExGT{t}{e}{ \ind{\tau\in du} \ind{\tau_I<\tau_C+\delta} \phi_u } = \ind{\tau>t} du\,\ExFT{t}{e}{\lambda^{\delta,I}_u D(t,u;\lambda) \phi_u }
\]%
where $\phi_u$ is ${\cal F}$-adapted and measurable at $u$. Furthermore, we introduce the pre-default intensities of each name modified to take into account the possibility that the surviving name defaults within the margin period of risk
\begin{eqnarray}
\label{eq:lambda_modified}
\ind{\tau>t} \lambda^{\delta,C}_t &:=& \ind{\tau>t} \lambda^C_t + \ind{\tau>t} \lambda^I_t \left( 1 - D(t,t+\delta;\lambda^C) \right) \\\nonumber
\ind{\tau>t} \lambda^{\delta,I}_t &:=& \ind{\tau>t} \lambda^I_t + \ind{\tau>t} \lambda^C_t \left( 1 - D(t,t+\delta;\lambda^I) \right)
\end{eqnarray}%
where we define the pre-default intensity $\lambda_t^I$ of the clearing member and the pre-default intensity $\lambda_t^C$ of the client as
\begin{eqnarray}
\label{eq:lambda}
\ind{\tau_I>t} \lambda_t^I \,dt &:=& \QxCT{e}{\tau_I\in dt}{\tau_I>t,{\cal F}_t} \\\nonumber
\ind{\tau_C>t} \lambda_t^C \,dt &:=& \QxCT{e}{\tau_C\in dt}{\tau_C>t,{\cal F}_t}
\end{eqnarray}%
along with their sum $\lambda_t$
\begin{equation}
\label{eq:lambda_sum}
\ind{\tau>t} \lambda_t := \ind{\tau>t} \lambda_t^I + \ind{\tau>t} \lambda_t^C
\end{equation}%
and we omit the tilde over the intensity symbols to lighten the notation.

\subsection{The Pricing Equation}
\label{sec:pe}

If we consider interest-rate derivatives we can apply the above results and switch to the default-free market filtration ${\cal F}_t$. We can do that since the defaultable names are only the clearing member and the client, so that when a default occurs the credit-risk-free close-out amount is ${\cal F}$-adapted.

A different situation occurs for credit derivatives. In this case the close-out amount is fully ${\cal G}$-adapted since the reference names of the credit derivatives may survive the default event and we must consider them even in a credit-risk-free close-out evaluation. We address the readers to \cite{BrigoCapponiPallavicini} for a discussion of the case of credit default swaps.

In the following section we consider the case of an ${\cal F}$-adapted close-out, and we derive the pricing equations for interest-rate derivatives.

\subsubsection{Switching to the Default-Free Market Filtration}
\label{sec:market_filtration}

Hence, since the pre-default processes are uniquely defined by their values up to the default event, we switch to the default-free market filtration ${\cal F}_t$, and we obtain
\begin{eqnarray}
\label{eq:fundingexplicitF}
{\widetilde V}_t
& = & \varepsilon_t(t,T) \\\nonumber
& + & \int_t^T du \,\ExFT{t}{e}{ D(t,u;f+\lambda^C+\lambda^I) ( \ell^{N^C}_u N^C_u + \ell^{N^I}_u N^I_u ) } \\\nonumber
& - & \int_t^T du \,\ExFT{t}{e}{ \lambda^{\delta,C}_u \,\lgd_C \,D(t,u;f+\lambda^C+\lambda^I) (\varepsilon_{u+\delta}(u,T) - \varepsilon_u(u,T) - N^C_u)^+ } \\\nonumber
& - & \int_t^T du \,\ExFT{t}{e}{ \lambda^{\delta,I}_u \,\lgd_I \,D(t,u;f+\lambda^C+\lambda^I) (\varepsilon_{u+\delta}(u,T) - \varepsilon_u(u,T) - N^I_u)^- }
\end{eqnarray}%
with
\[
\varepsilon_s(t,T) = \int_t^T \ExFT{s}{e}{ D(t,u;e) \Pi(u,u+du) }
\]%
where the initial margin posted by the counterparty (or the client in the CCP case) is defined as
\[
N^C_t = \Phi^{-1}_{0,\nu_t}(q)
\;,\quad
\nu^2_t = \VxFT{t}{h}{\varepsilon_{t+\delta}(t,T)}
\]%
while the initial margin posted by the investor in bilateral contracts under CSA is
\[
N^I_t = - N^C_t
\]
being zero in the CCP case. Moreover, we have
\[
\lambda^{\delta,C}_t = \lambda^C_t + \lambda^I_t \left( 1 - D(t,t+\delta;\lambda^C) \right)
\;,\quad
\lambda^{\delta,I}_t = \lambda^I_t + \lambda^C_t \left( 1 - D(t,t+\delta;\lambda^I) \right)
\]
along with
\[
f_t = e_t + \ell^+_t \ind{{\widetilde V}_t>\varepsilon_t(t,T)} + \ell^-_t \ind{{\widetilde V}_t<\varepsilon_t(t,T)}
\]%
{ 
where the free parameters to be calibrated on the market are the overnight rate $e_t$, and the default intensities of the two names $\lambda^C_t$ and $\lambda^I_t$. The liquidity bases $\ell_t^+$, $\ell_t^-$, $\ell^{N^C}_t$ and $\ell^{N^I}_t$ are linked to market data by means of equations \eqref{eq:liquidity_basis_I} and \eqref{eq:liquidity_basis_C}, while the margin period of risk $\delta$ and the confidence level $q$ of the risk measure used to estimate the initial margin can be inferred from contract documentation.
}

The final result is a pricing equation where the cash flows of the derivative are discounted at the  collateral rate adjusted for the funding costs of the initial margin and for the CVA/DVA terms. In the following sections we introduce a dynamical model to simulate the overnight rate, along with the liquidity bases, and the default intensities. Notice that the choice of the liquidity bases appearing in equation \eqref{eq:fundingexplicitF} depends on which is the party performing the calculation. In the following we select a different set of bases according to the party.

Furthermore, since the funding rates depend on the derivative price, namely the same expectation we are trying to compute,  equation \eqref{eq:fundingexplicitF} cannot be used to explicitly evaluate the price of the contract as a straightforward expectation. This equation  should rather be interpreted as the solution of a BSDE with terminal condition at contract maturity $T$ as described in \cite{Ma2002} and similarly to what we have seen earlier, with (\ref{eq:fundingwithnettingset}) becoming the BSDE (\ref{eq:bsde}). 
 In the next section we describe the numerical scheme used to solve the BSDE.

\subsubsection{Numerical Scheme for the BSDE}
\label{sec:bsde_numerical}

We calculate the prices by means equation \eqref{eq:fundingexplicitF}. We solve the equation in terms of the following implicit iterative problem defined on the time grid $\{t = t_0,\ldots,t_i,\ldots,t_n=T\}$ as done in \cite{Perini2011}
\[
Y_{t_i} = \ExFT{t_i}{e}{Y_{t_{i+1}} D(t_i,t_{i+1};f+\lambda^C+\lambda^I) } + \int_{t_i}^{t_{i+1}} \ExFT{t_i}{e}{ D(t_i,u;f+\lambda^C+\lambda^I) \,d\pi_u }
\]%
with terminal condition
\[
Y_{t_n} = 0
\]%
where we define the price process $Y_t$ as
\begin{equation}
\label{eq:Y}
Y_t := {\widetilde V}_t - \varepsilon_{t}(t,T)
\end{equation}%
and the coupon process $\pi_t$ as given by
\begin{eqnarray}
\label{eq:pi}
d\pi_t
&:=& \ell^{N^C}_t N^C_t \,dt - \lambda^{\delta,C}_t \,\ExFT{t}{e}{ \lgd_C (\varepsilon_{t+\delta}(t,T) - \varepsilon_t(t,T) - N^C_t)^+ } \,dt \\\nonumber
& + & \ell^{N^I}_t N^I_t \,dt - \lambda^{\delta,I}_t \,\ExFT{t}{e}{ \lgd_I (\varepsilon_{t+\delta}(t,T) - \varepsilon_t(t,T) - N^I_t)^- } \,dt
\end{eqnarray}%
It is important that the fundamental nonlinearity induced by the funding problem has disappeared from the coupon process $\pi$, since this does not depend on rates $f$ for example, and as such does not depend on the future values of the very same value $V$ we are trying to compute. See Remark \ref{rem:nonlin}. This important feature of our coupon process is a byproduct of our specific choices for $\varepsilon$, $M$ and $\ell$ in the interest rate case and on our simplifying assumptions on the treasury netting sets for funding. The remaining presence of $f$ in the discount term will be dealt with numerically now.

In order to implement numerically the problem we build a Euler discretization as in \cite{Ma2002}, and we write
\[
Y_{t_i} = \left(1 - g_{t_i}(Y_{t_i}) \Delta t_i  \right) \ExFT{t_i}{e}{ Y_{t_{i+1}} } + \Delta\pi_{t_i}
\]%
where we make explicit the dependency on the mark-to-market value of the contract and we define the rate
\[
g_{t_i}(Y_{t_i}) := e_{t_i} + \lambda^C_{t_i} + \lambda^I_{t_i} + \ell^+_{t_i} \ind{Y_{t_i}>0} + \ell^-_{t_i} \ind{Y_{t_i}<0} \,.
\]%

Then, we follow \cite{Ma2002} by solving the coupled pair of equations by a fixed-point technique, leading to the following explicit scheme
\begin{equation}
\label{eq:explicit}
Y_{t_i} = \left(1 - g_{t_i}\left(\ExFT{t_i}{e}{ Y_{t_{i+1}} }\right) \Delta t_i \right) \ExFT{t_i}{e}{ Y_{t_{i+1}} } + \Delta\pi_{t_i} \,.
\end{equation}%

We can reduce the variance of the simulation by means of control variate variables based on the price of the contract without funding costs. We define the funding-cost-free pre-default price as given by
\begin{eqnarray}
\label{eq:fcf_price}
{\widetilde V}^0_t
& := & \varepsilon_t(t,T) \\\nonumber
& + & \int_t^T du \,\ExFT{t}{e}{ D(t,u;e+\lambda^C+\lambda^I) ( \ell^{N^C}_u N^C_u + \ell^{N^I}_u N^I_u ) } \\\nonumber
& - & \int_t^T du \,\ExFT{t}{e}{ \lambda^{\delta,C}_u \,\lgd_C \,D(t,u;e+\lambda^C+\lambda^I) (\varepsilon_{u+\delta}(u,T) - \varepsilon_u(u,T) - N^C_u)^+ } \\\nonumber
& - & \int_t^T du \,\ExFT{t}{e}{ \lambda^{\delta,I}_u \,\lgd_I \,D(t,u;e+\lambda^C+\lambda^I) (\varepsilon_{u+\delta}(u,T) - \varepsilon_u(u,T) - N^I_u)^- }
\end{eqnarray}%
This price process can be evaluated directly by calculating the expectation, since the right-hand side does not depend on ${\widetilde V}^0_t$. The problem we had pointed out in Remark \ref{rem:nonlin} does not hold for this price process. This is basically because $\pi$ is free from future values of $V$ and now also the discounting, having been replaced by $e$, is free from future $V$'s.
Yet, we can evaluate this price also by means of the numerical scheme of equation \eqref{eq:explicit}. If we introduce the funding-cost-free price process $Y^0_t$ as
\[
Y^0_t := {\widetilde V}^0_t - \varepsilon_{t}(t,T) \,,
\]%
we get
\[
Y^0_{t_i} = \left(1 - (e_{t_i} + \lambda^C_{t_i} + \lambda^I_{t_i}) \Delta t_i \right) \ExFT{t_i}{e}{ Y^0_{t_{i+1}} } + \Delta\pi_{t_i} \,.
\]

We can solve for the coupon process $\pi_{t_i}$ the above equation, and substitute the resulting expression into equation \eqref{eq:explicit}, to obtain a numerical scheme to evaluate the funding valuation adjustment process $X_t := Y_t - Y^0_t$. We obtain
\begin{equation}
\label{eq:explicit_cv}
\left\{
\begin{aligned}
& {\cal X}_{t_i} := \ExT{t_i}{e}{X_{t_{i+1}}} \\
& {\cal Y}_{t_i} := {\widetilde V}^0_{t_i} - \varepsilon_{t_i}(t_i,T) + {\cal X}_{t_i} \\
& X_{t_i} = {\cal X}_{t_i} - {\cal Y}_{t_i} \,g_{t_i}\!\left({\cal Y}_{t_i}\right) \Delta t_i
\end{aligned}
\right.
\end{equation}%
with terminal condition $X_{t_n} = 0$. We can solve iteratively the scheme, and we obtain the present value of the funding valuation adjustment process, namely $X_{t_0}$, which allows to calculate the derivative price as
\begin{equation}
\label{eq:wfc_price}
{\widetilde V}_{t_0} = {\widetilde V}^0_{t_0} + X_{t_0} \,.
\end{equation}%

\subsection{Price Decomposition: CVA, DVA, FVA, and MVA}
\label{sec:price_decomp}

In our previous works we pointed out that the funding inclusive price, in general, is not really separable in clearcut components depending only on credit risk, debit risk and funding risk, see for example \cite{Perini2011,Perini2012}. {  However, under the simplifying assumptions we adopted here in the interest rate case, we have now approached a specific situation where separability can be partially achieved.}

Indeed, the control-variate procedure highlights a simple way to decompose the derivative price in five pieces with financial meaning. We re-write equation \eqref{eq:wfc_price} by plugging into it equation \eqref{eq:fcf_price} to discuss the terms. We obtain
\begin{equation}
\label{eq:price_decomp}
{\widetilde V}_t = \mtm_t + \cva_t + \dva_t + \mva_t + \fva_t
\end{equation}%
where the five terms are
\begin{enumerate}
\item The mark-to-market of the replacement deal used to evaluate the close-out amount, which in our settings corresponds to the price of the contract in the hypothetical settings of complete "perfect"  collateralization (no gap risk, continuous updating), namely
\[
\mtm_t := \varepsilon_t(t,T) \,.
\]%
\item {  The funding-cost-free bilateral CVA adjustment}, which is due to the default of the client (''C``), and it is corrected for the possibility of default of the clearing member during the margin period of risk. The adjustment is equal to
\[
\cva_t := - \int_t^T du \,\ExFT{t}{e}{ \lambda^{\delta,C}_u \,\lgd_C \,D(t,u;e+\lambda^C+\lambda^I) (\varepsilon_{u+\delta}(u,T) - \varepsilon_u(u,T) - N^C_u)^+ } \,.
\]%
\item {  The funding-cost-free bilateral DVA adjustment}, which is due to the default of the clearing member (''I``), and it is corrected for the possibility of default of the client during the margin period of risk. The adjustment is equal to
\[
\dva_t := - \int_t^T du \,\ExFT{t}{e}{ \lambda^{\delta,I}_u \,\lgd_I \,D(t,u;e+\lambda^C+\lambda^I) (\varepsilon_{u+\delta}(u,T) - \varepsilon_u(u,T) - N^I_u)^- } \,.
\]%
\item The funding costs of posting the initial margin. We can call them the margin valuation adjustment, or MVA, and we define it as
\[
\mva_t := \int_t^T du \,\ExFT{t}{e}{ \,D(t,u;e+\lambda^C+\lambda^I) ( N^C_u \ell^{N^C}_u + N^I_u \ell^{N^I}_u ) } \,.
\]%
\item The funding costs to implement the hedging strategy by taking into account the cash obtained from the re-hypothecation of the variation margin. {  This term {\it cannot} be directly calculated} by means of an expectation, but it requires the numerical solution of a BSDE. It is given by
\[
\fva_t := X_t
\]%
where $X_t$ is calculated by means of the numerical iterative scheme \eqref{eq:explicit_cv}.
\end{enumerate}

We should point out that in some contexts it may make sense to include MVA into FVA, as the cost of margining is strictly speaking a funding cost. We maintained distinction between the two for clarity.

The above decomposition is valid either for CCP cleared contracts or for bilateral trades, also under Standard CSA with the relevant definitions of initial margin accounts and liquidity bases. The only limitation is given by the selection of an ${\cal F}$-adapted close-out amount, and the ${\cal F}$-conditional independence of default times. When one of these requirements is not met, we need to use directly equation \eqref{eq:fundingexplicit} to price the derivative, without switching to the default free market filtration.

\section{Numerical Results and Discussion}
\label{sec:numerics}

In this section we apply the pricing equation \eqref{eq:fundingexplicitF} to the case of an interest-rate swap (IRS) cleared by a CCP. In particular, we adopt the numerical scheme (\ref{eq:explicit_cv}), and we analyze receiver and payer IRS with a maturity of ten years with Euro market conventions. Our goal is studying the impact of the liquidity basis and of the correlation between market and credit risks (wrong-way risk).

\subsection{Modelling the Market and Default Risks}
\label{sec:model}

The pricing equation \eqref{eq:fundingexplicitF} depends on the interest-rates, the liquidity basis, and the default intensities. In this numerical section to simplify the discussion we work in a single-curve framework. However, the generalization to a multiple-curve setup is straightforward and can be accomplished, for instance, by following the model presented in \cite{Moreni2010,Moreni2012} and used later in \cite{BrigoPallavicini2013}.

\begin{figure}
\begin{center}
\includegraphics[width=7.5cm, height = 6cm]{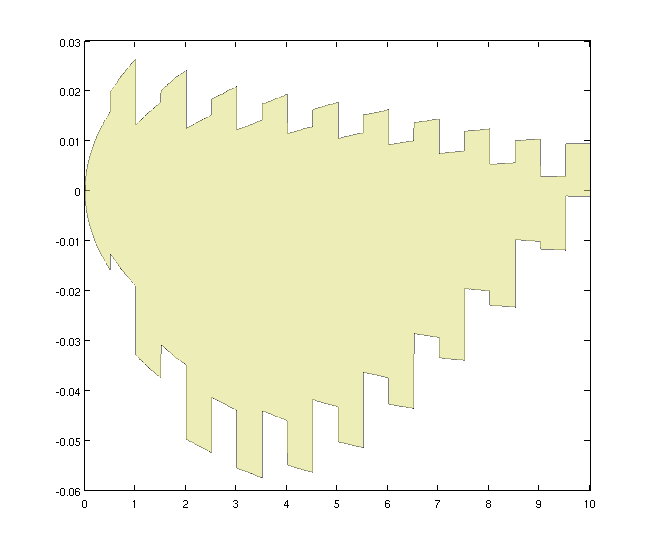}
\end{center}
\caption{Positive and negative exposures of collateralized ten-year receiver IRS traded at par, which we use as example in the numerical discussion. Collateral variation margin matches the mark to market at the discrete margining dates. There is no instantaneous gap risk. We still have margin period gap risk. The notional is one Euro and prices in basis points.}
\label{fig:exposure}
\end{figure}

\begin{figure}
\begin{center}
\includegraphics[width=7.5cm, height = 6cm]{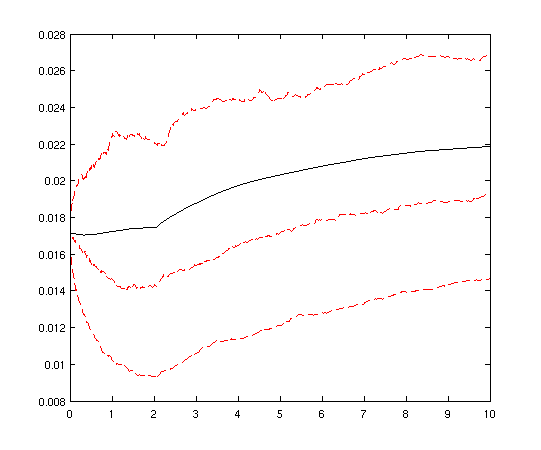}
\includegraphics[width=7.5cm, height = 6cm]{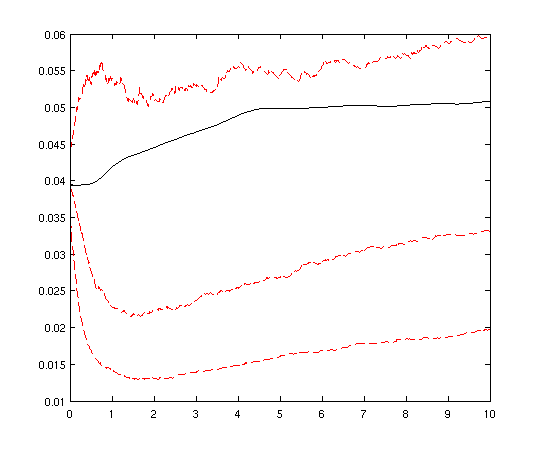}
\end{center}
\caption{Left: default intensity in ''M`` market setting as a zero rate (solid line) along with distribution quantiles at 0.25, 0.5 and 0.75 level. Right: default intensity in ''H`` market setting as in central panel. See market data settings in the appendix.}
\label{fig:intensitiesperc}
\end{figure}

We model interest-rates under the pricing measure of equation \eqref{eq:fundingexplicitF} as in \cite{BrigoPallaviciniPapatheodorou} as a two-factor shifted Hull-White model, calibrated to swaption volatilities. Thus, we can write
\[
e_t := \varphi(t) + \sum_{i=1}^2 x^i_t
\]%
where $\varphi(t)$ is a deterministic function of time used to calibrate the initial term structure of zero-coupon rates, and
\begin{equation}
\label{eq:hw}
dx^i_t = -a^i x^i_t \,dt + \sigma^i \,dW^i_t \;,\quad x^i_0=0
\end{equation}%
where $a^i$ and $\sigma^i$ are positive constants, and $W^i_t$'s are standard Brownian motions. Since we are working in a single-curve framework we can derive all the other market rates from the overnight rate. Calibration details can be found in \cite{BrigoPallaviciniPapatheodorou}, while calibration data are reported in Tables \ref{disc} and \ref{swapvol}. Here, we show in  Figure \ref{fig:exposure} the resulting exposures for a collateralized ten-year IRS traded at par, with collateral matching mark to market on margining dates, which we use as example in the numerical discussion.

We model the default intensities according to a one-factor shifted Cox-Ingersoll-Ross model as given by
\[
\lambda^k_t := y^k_t + \psi^k(t)
\;,\quad
k\in\{I,C\}
\]%
where $\psi^k(t)$ is a deterministic function of time used to calibrate the initial term structure of credit spreads, and
\begin{equation}
\label{eq:cir}
dy^k_t = \kappa^k (\mu^k-y^k_t) \,dt + \nu^k \sqrt{y^k_t} \,dW^k_t
\end{equation}%
{ 
where $\kappa^k$, $\mu^k$, $\nu^k$ and $y^k_0$ are positive constants, and $W^k_t$ is a standard Brownian motion. Since we do not have a liquid market for CDS options, we use as reference calibration set the initial term structure and the options used in \cite{BrigoPallaviciniPapatheodorou} to calibrate default intensities. Calibration data are reported in Tables \ref{tab:cdstermmid} and \ref{tab:cdsvolsmid} for the ''mid-risk`` (or ''M``) market settings, while in the right panel of figure \ref{fig:intensitiesperc} we illustrate the resulting default-intensity distribution. In Tables \ref{tab:cdstermhigh} and \ref{tab:cdsvolshigh} and in the left panel of figure \ref{fig:intensitiesperc} we report the same data for the ''high-risk`` (or ''H``) market settings.
}

The four Brownian motions are correlated by means of a correlation matrix with elements $\rho^{ij}$ defined as
\[
d \langle W^i, W^j\rangle_t = \rho^{ij} \,dt
\]%
It is useful to calculate the correlation among the overnight rate and each default intensity, since we use it in the numerical discussion. We get
\begin{equation}
\label{eq:rho}
\bar{\rho}^k := \frac{d \langle e , \lambda^k \rangle_t}{\sqrt{d \langle e , e \rangle_t\ d \langle \lambda^k , \lambda^k \rangle_t}} 
= \frac{\sum_{i=1}^2 \sigma^i \rho^{ik}}{\sqrt{\sum_{i=1}^2\sum_{j=1}^2 \sigma^i \sigma^j \rho^{ij}}}
\end{equation}%

In the following sections we price a ten-year interest-rate swap in different credit and liquidity settings. We start from a bilateral trade without initial margin and with initial margins as in SCSA. Then we focus on CCP cleared contracts. The calculations are made with the made settings of \cite{BrigoCapponiPallaviciniPapatheodorou}, so that prices without funding costs can be found on that paper, which we reproduce in the appendix.

\subsection{Interest-Rate Swap: bilateral trades without margining}

In this section we investigate the impact of funding costs in a bilateral trade without collateralization, as in deals a bank may close with a corporate counterparty without a CSA. The payoff we discuss is a ten year interest-rate swap, which we analyze both in the receiver and payer variants. The strike of these swaps is chosen so that the price is zero when credit and liquidity adjustments are disregarded (full collateralization case). The recovery rates for both parties in the trade are set to $40\%$.

\begin{figure}
\begin{center}
\includegraphics[width=7.5cm, height = 6cm]{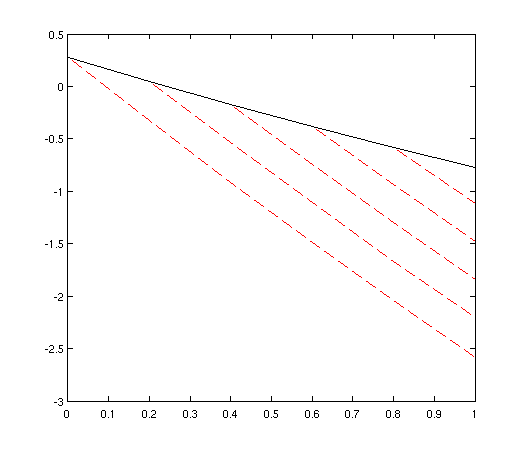}
\includegraphics[width=7.5cm, height = 6cm]{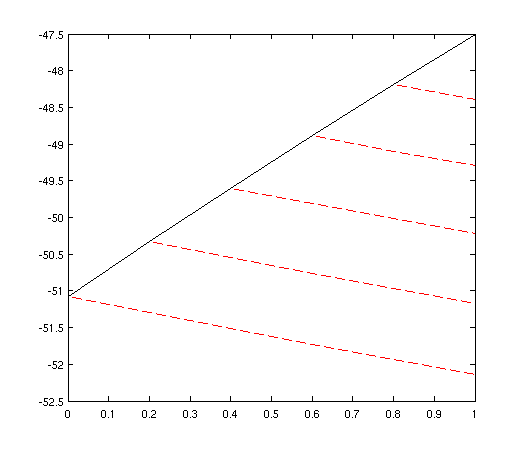}
\end{center}
\caption{Prices of a ten-year receiver IRS (left) and payer IRS (right) with a high-risk parameter set for the counterparty and a mid-risk parameter set for the investor (''H/M``). In each panel the black continuous line represents a funding policy with the same rate for funding and investing ($\beta^+=\beta^-$), while the dashed red lines represent funding policies with asymmetric rates ($\beta^+\geq\beta^-$) for different choices of the investing rate. The $x$-axis shows the funding multiplier $\beta^+$. For each red plot, the multiplier $\beta^-$ is taken equal to the value of $\beta^+$ at the intersection with the black line. Prices are inclusive of CVA and DVA and they are expressed in basis points with a notional of one Euro.}
\label{fig:funding_policy_HM}
\end{figure}

We start by pricing our reference IRS under different funding policies by assuming that there is no correlation among credit and market risks, namely that $\bar{\rho}^C=\bar{\rho}^I=0$ in equation \eqref{eq:rho}. In particular, we change the liquidity multipliers of equations \eqref{eq:liquidity_basis_I} and \eqref{eq:liquidity_basis_C} to simulate the presence of different levels of funding fees along with the possibility of asymmetric rates. In Figure \ref{fig:funding_policy_HM} we show the results with high-risk parameter set for the counterparty and mid-risk parameter set for the investor (''H/M``). See the appendix for full details on market data. 

\begin{figure}
\begin{center}
\includegraphics[width=7.5cm, height = 6cm]{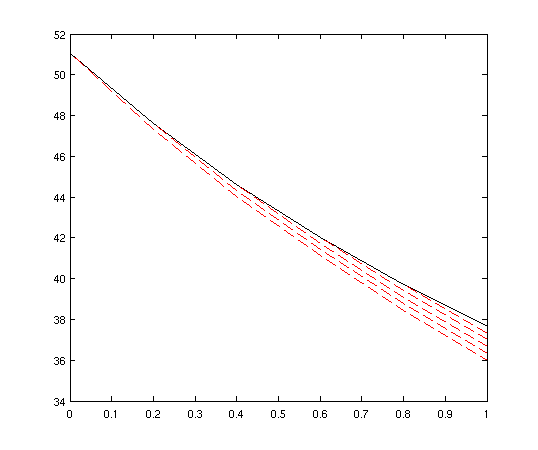}
\includegraphics[width=7.5cm, height = 6cm]{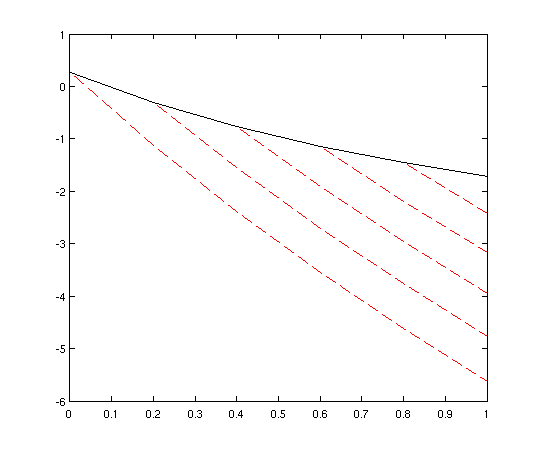}
\end{center}
\caption{Prices of a ten-year receiver IRS (left) and payer IRS (right) with a mid-risk parameter set for the counterparty and a high-risk parameter set for the investor (''M/H``). All else as in the caption of Figure \ref{fig:funding_policy_HM}.}
\label{fig:funding_policy_MH}
\end{figure}

The impact of funding rates seems to show a roughly linear pattern with respect to the value of the funding and investing rates for the market settings used for the calculations. However, if we extend our analysis to different settings we can observe more complex patterns. In Figure \ref{fig:funding_policy_MH} we show the results with mid-risk parameter set for the counterparty and high-risk parameter set for the investor (''M/H``). In this way the funding costs are higher and non-linear patterns appear. 

\begin{figure}
\begin{center}
\includegraphics[width=7.5cm, height = 6cm]{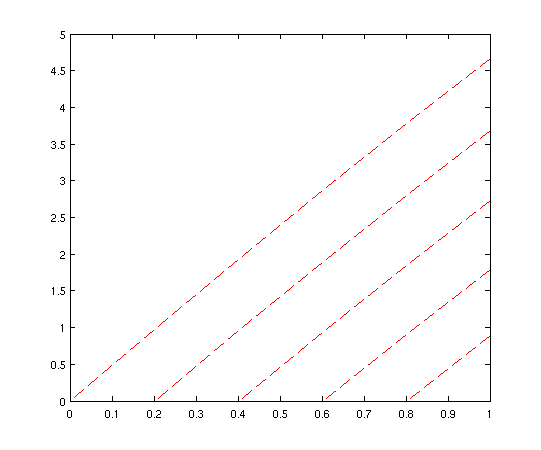}
\includegraphics[width=7.5cm, height = 6cm]{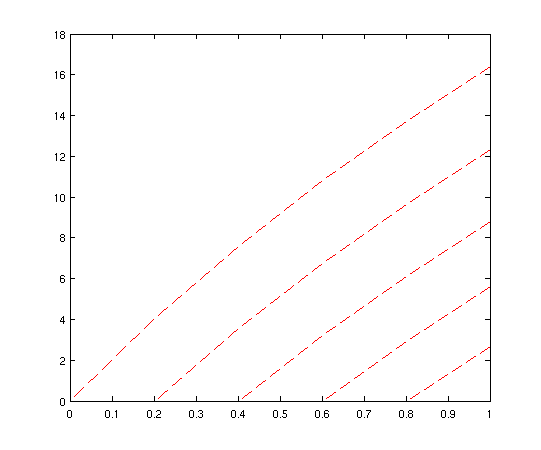}
\end{center}
\caption{Bid-ask spread of a ten-year receiver IRS with a high-risk parameter set for the counterparty and a mid-risk parameter set for the investor (''H/M``) on the left, and the opposite situation (''M/H``) on the right. In each panel the dashed red lines represent funding policies with asymmetric rates ($\beta^+\geq\beta^-$) for different choices of the investing rate, that is set at the intersection of the red line with the $x$ axis. The $x$-axis shows the funding multiplier $\beta^+$. Prices are inclusive of CVA and DVA and they are expressed in basis points with a notional of one Euro.}
\label{fig:differential_receiver}
\end{figure}

In presence of asymmetric rates, namely when the funding (borrowing) rate is greater than the investing (lending) rate, we have that the price of a long position plus the price of the analogous short position  is not identically zero, since the pricing operator is not linear in this case, leading to differences between the price of a portfolio and the portfolio of the prices. Such differences may be interpreted, possibly, as the bid-ask spread for the position. See \cite{Bergman1995} or \cite{Mercurio2013} for a discussion. We observe such effects in Figure \ref{fig:differential_receiver} where we calculate the sum of the price of the receiver position plus the price of shorting the same positions for the ''H/M`` and ''M/H`` market settings respectively. The resulting bid-ask spread is zero when the funding rate is equal to the investing rate, while it grows as soon as the difference between these two rates increases.

\begin{figure}
\begin{center}
\includegraphics[width=7.5cm, height = 6cm]{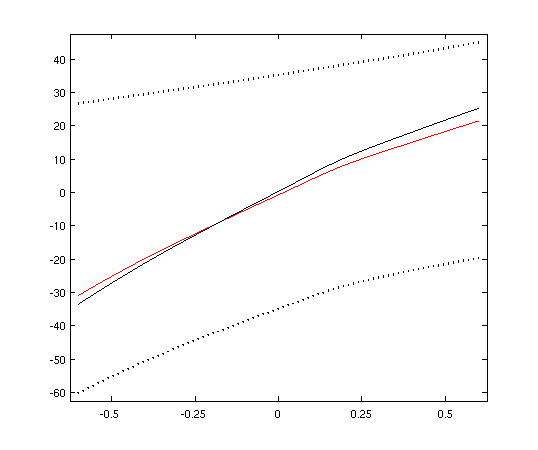}
\includegraphics[width=7.5cm, height = 6cm]{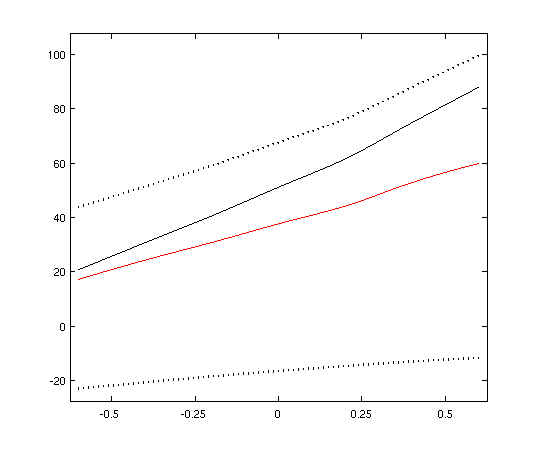}
\end{center}
\caption{Prices of a ten-year receiver IRS with a high-risk parameter set for the counterparty and a mid-risk parameter set for the investor (''H/M``) on the left, and the opposite situation (''M/H``) on the right. In each panel the black continuous line represents the price inclusive of CVA and DVA but not funding costs, with the dashed black lines representing separately the CVA and the DVA contributions. The red continuous line is the price inclusive both of credit and funding costs. Symmetric funding policy with $\beta^+=\beta^-=1$. On the $x$-axis the correlation among market and credit risks ($\bar{\rho} = \bar{\rho}^C = \bar{\rho}^I$). Prices in basis points with a notional of one Euro.}
\label{fig:wwr_funding_receiver}
\end{figure}

Now, we can move onto analyzing the impact of correlations among market and credit risks. We consider a receiver payoff both in ''H/M`` and ''M/H`` market settings. We maintain as funding policy $\beta^+=\beta^-=1$. The results are shown in Figure \ref{fig:wwr_funding_receiver}. We observe two interesting features: (i) funding costs may be dominated by funding benefits (left panel for negative correlations), and (ii) funding costs may be a relevant part of credit/liquidity adjustments (right panel for positive correlations).

Detailed sets of prices for the market setting ''H/M'' for different funding policies and different correlation among market and credit risks are listed in Tables \ref{tab:irs_rec_sym_wwr}, \ref{tab:irs_rec_asym_wwr}, \ref{tab:irs_pay_sym_wwr}, \ref{tab:irs_pay_asym_wwr}, and \ref{tab:irs_rec_diff} in the appendix.

\subsection{Interest-Rate Swap: bilateral CSA trades with margining}

We move onto analyzing collateralization. We still consider a bilateral trade but under a CSA. In particular, we discuss the setting presented in equation \eqref{eq:fundingexplicitF} with the possibility of an incomplete collateralization, as discussed in Remark \ref{rem:vmfrac}, when the initial margin is not present. Detailed numerical results can be found in Tables \ref{tab:irs_rec_sym_vm_wwr} and \ref{tab:irs_pay_sym_vm_wwr}. The recovery rates for both parties in the trade are set to $40\%$.

We wish to compare the effectiveness of a collateralization based only on variation margins with a collateralization based also on posting the initial margin. In the former case we should face residual CVA/DVA effects coming either from an incomplete implementation of the margining procedure or from a margin period of risk (we consider ten days in these examples), while in the second case we have to pay for the costs of funding the initial margin.

\begin{figure}
\begin{center}
\includegraphics[width=7.5cm, height = 6cm]{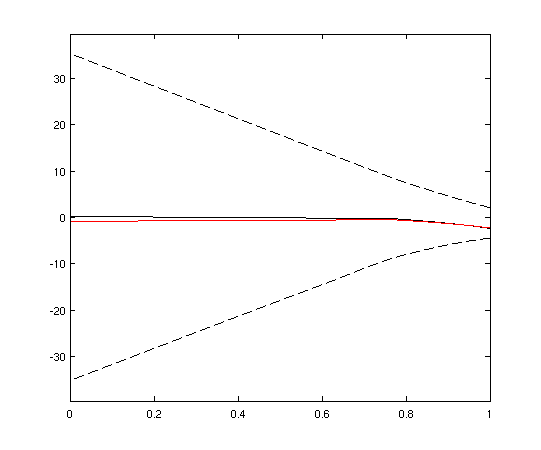}
\includegraphics[width=7.5cm, height = 6cm]{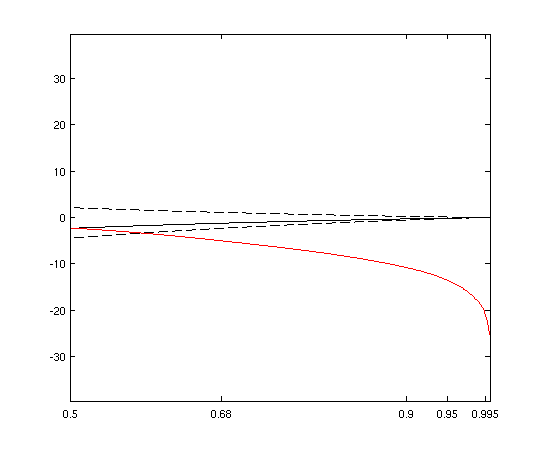}
\end{center}
\caption{Prices of a ten-year receiver IRS with a high-risk parameter set for the counterparty and a mid-risk parameter set for the investor (''H/M``). On the left prices for different collateralization fractions $\alpha$ ($x$-axis) without initial margin. On the right prices with $\alpha=1$ and initial margin posted at various confidence levels $q$ ($x$-axis). In each panel the black continuous line represents the price inclusive of CVA and DVA but not funding costs, with the dashed black lines representing separately the CVA and the DVA contributions. The red continuous line is the price inclusive both of credit and funding costs. Symmetric funding policy with $\beta^+=\beta^-=1$. Correlation $\bar\rho$ is zero. Prices in basis points with a notional of one Euro.}
\label{fig:irs10y_csa_receiver_HM}
\end{figure}

\begin{figure}
\begin{center}
\includegraphics[width=7.5cm, height = 6cm]{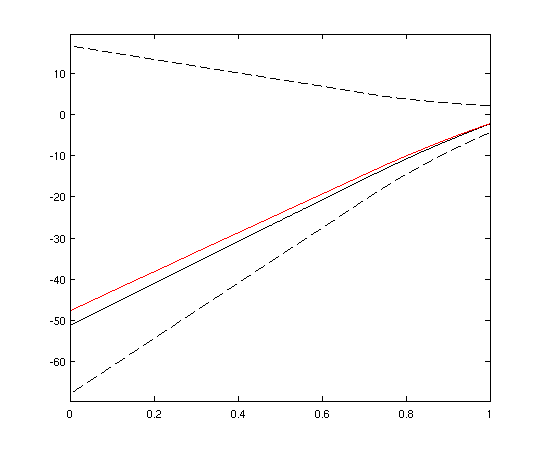}
\includegraphics[width=7.5cm, height = 6cm]{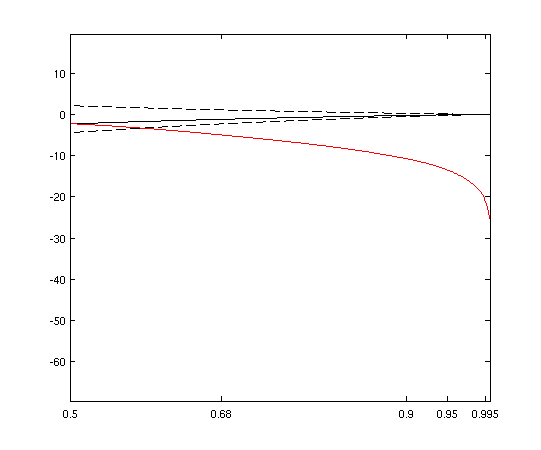}
\end{center}
\caption{Prices of a ten-year \emph{payer} IRS with a high-risk parameter set for the counterparty and a mid-risk parameter set for the investor (''H/M``). All else as in the caption of 
Figure~\ref{fig:irs10y_csa_receiver_HM}
}
\label{fig:irs10y_csa_payer_HM}
\end{figure}

\begin{figure}
\begin{center}
\includegraphics[width=7.5cm, height = 6cm]{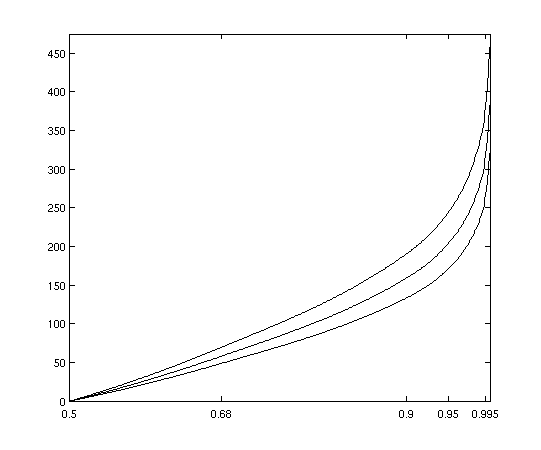}
\includegraphics[width=7.5cm, height = 6cm]{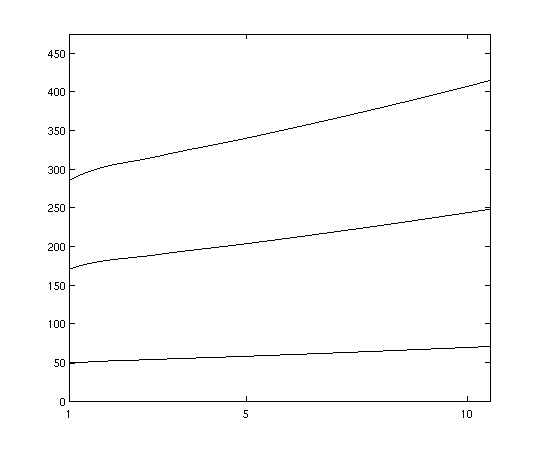}
\end{center}
\caption{Amount of initial margin requested at contract inception for a ten-year receiver IRS with a high-risk parameter set for the counterparty and a mid-risk parameter set for the investor (''H/M``). On the left the $x$-axis lists different confidence levels, while the curves correspond to three different margin period of risks ($1$, $5$ and $10$ days). On the right the $x$-axis lists different margin period of risks, while the curves correspond to three confidence levels ($68\%$, $95\%$ and $99.7\%$). Symmetric funding policy with $\beta^+=\beta^-=1$. Correlation $\bar\rho$ is zero. Initial margin amounts are expressed in basis points with a notional of one Euro.}
\label{fig:irs10y_im_receiver_HM}
\end{figure}

In Figure \ref{fig:irs10y_csa_receiver_HM} we show the prices for a receiver IRS in the ``H/M'' market settings, while in Figure \ref{fig:irs10y_csa_payer_HM} the payer case is presented. If we consider equations \eqref{eq:initial_margin_C} and \eqref{eq:initial_margin_I}, or their approximations \eqref{eq:initial_margin_approx_C} and \eqref{eq:initial_margin_approx_I}, we can see that the initial margin is used to cover extreme events due to the default either of the investor or the counterparty. The quantity of initial margin required to cover such extreme events at high confidence levels, as requested by ISDA or by CCPs, may result in a sensitive amount of funding costs. As an example we show in Figure \ref{fig:irs10y_im_receiver_HM} the amount of initial margin requested at inception for a receiver IRS in the ``H/M'' market settings.

\subsection{Interest-Rate Swap: CCP cleared contract}

Our last example is on CCP cleared contracts. We focus on the client side and we consider the amount of funding costs the client is paying to trade with the CCP. The recovery rate for the client is set to $40\%$, while the recovery rate for the clearing member is set to $95\%$, so as to model a possible loss resulting from the auction needed to move the trade to the backup member in case of clearing member default.

\begin{figure}
\begin{center}
\includegraphics[width=7.5cm, height = 6cm]{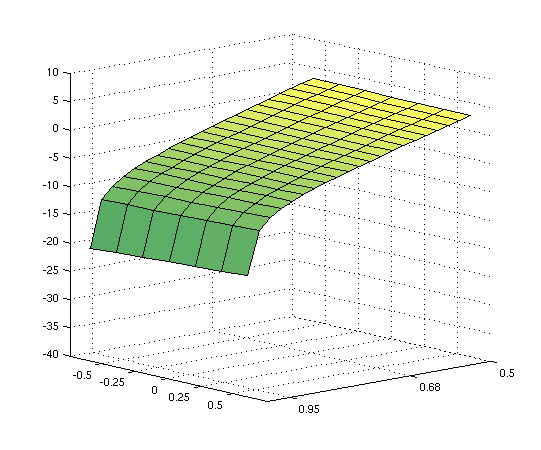}
\includegraphics[width=7.5cm, height = 6cm]{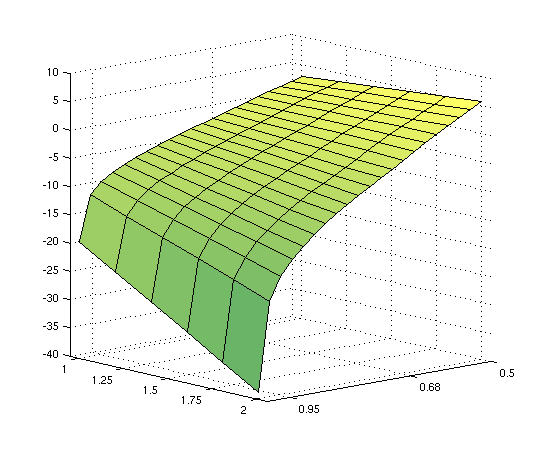}
\end{center}
\caption{Prices of a ten-year receiver IRS traded with a CCP from the point of view of the client with a mid-risk parameter set for the clearing member and a high-risk parameter set for the client for initial margin posted at various confidence levels $q$. On the left we display different correlations among market and credit risks ($\bar{\rho} = \bar{\rho}^C = \bar{\rho}^I$). On the right different levels of the volatility multiplier for interest rates. Symmetric funding policy with $\beta^+=\beta^-=1$. Prices in basis points with a notional of one Euro.}
\label{fig:irs10y_ccp_MH}
\end{figure}

Firstly, we show in Figure \ref{fig:irs10y_ccp_MH} the funding costs for different confidence levels by varying either the correlation among market and credit risks or the volatility of interest rates. We vary the volatility of interest rates by multiplying the volatilities $\sigma^1$ and $\sigma^2$ of equation \eqref{eq:hw} by a positive constant, the ``volatility multiplier". The margin period of risk is  five days. We notice a very little dependence on the market/credit correlation, while modifying the interest-rate volatilities has a much greater impact since the initial margin depends on the conditional variance of the remaining cash flows of the contract.

\begin{figure}
\begin{center}
\includegraphics[width=7.5cm, height = 6cm]{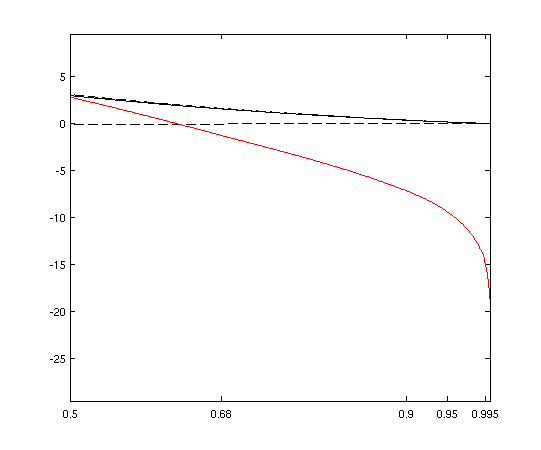}
\includegraphics[width=7.5cm, height = 6cm]{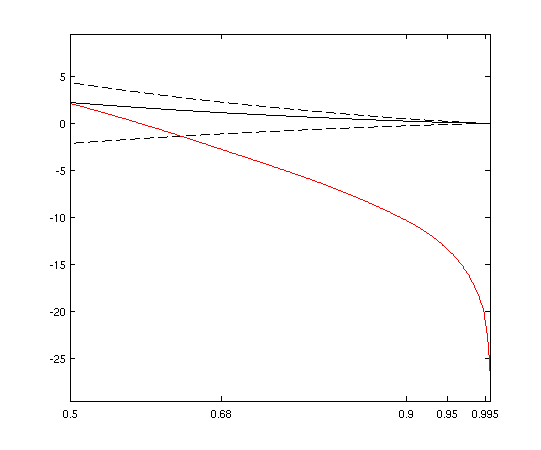}
\end{center}
\caption{Prices of a ten-year receiver IRS traded with a CCP (or bilaterally) with a mid-risk parameter set for the clearing member (investor) and a high-risk parameter set for the client (counterparty) for initial margin posted at various confidence levels $q$. Prices are calculated from the point of view of the client (counterparty). On the left the CCP trade, on the right the bilateral trade. In each panel the black continuous line represents the price inclusive of CVA and DVA but not funding costs, with the dashed black lines representing separately the CVA and the DVA contributions. The red continuous line is the price inclusive both of credit and funding costs. Symmetric funding policy with $\beta^+=\beta^-=1$. Correlation $\bar\rho$ is zero. Prices in basis points with a notional of one Euro.}
\label{fig:irs10y_ccp_receiver_MH}
\end{figure}

A second experiment is to compare the same IRS contract cleared by the CCP or bilaterally traded under CSA. We stay from the point of view of the same party (the CCP client in the first case, the counterparty in the second case) and we assume a margin period of risk of five days in the CCP trade, and of ten days in the bilateral case (these are typical values for such scenarios). We assume that in both case the other party, namely the clearing member in the CCP trade and the investor in the bilateral CSA trade, has the same credit quality. The results are shown in Figure \ref{fig:irs10y_ccp_receiver_MH}, whereas the valuation adjustments are tabulated in Table~\ref{tab:irs_rec_sym_decomp}. 
We can see that in the bilateral case a residual CVA and DVA is present unless high confidence levels are used to calculate the initial margin, while in the CCP cleared case the CVA is practically null. Furthermore, in the bilateral case there is a greater amount of funding costs since a longer margin period of risk is used. 

\section{Conclusions and Further Developments}
\label{sec:conclusion}

This paper introduces a pricing framework that is based only on market observables and that consistently includes credit, collateral and funding effects in the pricing of a trade.  We discussed both the case of a bilateral CSA trade and the case of a trade cleared by a CCP. We analyzed the detailed impact of credit, collateral (both initial and variation margins) and funding costs in both cases, inclusive of margin period of risk, gap risk and wrong way risk. In particular, we analyzed the impact of funding costs on the initial margins required by the clearing house. The pricing problem is solved numerically in some relevant cases, in the interest rate swaps market, to show the impact of different ways of calculating the initial margin and to seize the effect of correlations among market risk factors on one side and credit risk and liquidity bases on the other side. A side benefit of our analysis is the characterization of a case where our valuation equation can be separated into different valuation components due to credit, funding and margining, leading to rigorous MVAs and FVAs on top of the usual CVAs and DVAs. The separability result does not hold in general, as we pointed out in earlier works. Further work includes a more realistic model for initial margins, the extension of our analysis to different asset classes and a more detailed study of the funding policies and rates applied by treasury departments in banks.

\section*{Acknowledgments}

We are grateful to Luca Coppola for useful insights on CCP mechanics, Mark Henrard and St\'ephan Cr\'epey for stimulating questions and discussions, Sergio Solinas for computer facilities.

\newpage 

\appendix
 
\section{Interest-Rate Market Data, CDS Terms Structures and Implied Volatilities}\label{app:mkdata}

We report in Table~\ref{disc} the yield curve term structure, and in Table~\ref{swapvol} swaption volatilities used to calibrate the interest-rate and credit-spread dynamics in section~\ref{sec:model}. We list also in Table~\ref{tab:cdstermmid} and~\ref{tab:cdstermhigh} the CDS term structures, and in Table~\ref{tab:cdsvolsmid} and~\ref{tab:cdsvolshigh} implied volatilities associated with the model parameters used for the credit-spread dynamics in section~\ref{sec:model}.

\begin{table}
\caption{
EUR zero-coupon continuously-compounded spot rates (ACT/360) observed on May, 26 2009.
}
\label{disc}
\begin{center}
{\small
\begin{tabular}{|cc|cc|cc|}
\hline
{\bf Date} & {\bf Rate} & {\bf Date} & {\bf Rate} & {\bf Date} & {\bf Rate} \\\hline
 27-May-09 &     1.15\% &  28-Dec-09 &     1.49\% &  29-May-17 &     3.40\% \\
 28-May-09 &     1.02\% &  28-Jan-10 &     1.53\% &  28-May-18 &     3.54\% \\
 29-May-09 &     0.98\% &  26-Feb-10 &     1.56\% &  28-May-19 &     3.66\% \\
 04-Jun-09 &     0.93\% &  29-Mar-10 &     1.59\% &  28-May-21 &     3.87\% \\
 11-Jun-09 &     0.92\% &  28-Apr-10 &     1.61\% &  28-May-24 &     4.09\% \\
 18-Jun-09 &     0.91\% &  28-May-10 &     1.63\% &  28-May-29 &     4.19\% \\
 29-Jun-09 &     0.91\% &  30-May-11 &     1.72\% &  29-May-34 &     4.07\% \\
 28-Jul-09 &     1.05\% &  28-May-12 &     2.13\% &  30-May-39 &     3.92\% \\
 28-Aug-09 &     1.26\% &  28-May-13 &     2.48\% &  & \\
 28-Sep-09 &     1.34\% &  28-May-14 &     2.78\% &  & \\
 28-Oct-09 &     1.41\% &  28-May-15 &     3.02\% &  & \\
 30-Nov-09 &     1.46\% &  30-May-16 &     3.23\% &  & \\\hline
\end{tabular}}
\end{center}
\end{table}

\begin{table}
\caption{
Market at-the-money swaption volatilities observed on May, 26 2009. Each column contains volatilities of swaptions of a given tenor $b$ for different expiries $t$.
}
\label{swapvol}
\begin{center}
{\small
\begin{tabular}{|c|ccccc|}\hline
{\bf $\boldsymbol t\downarrow$ / $\boldsymbol b\rightarrow$}   & {\bf 1y} & {\bf 2y} & {\bf 3y} & {\bf 4y} & {\bf 5y} \\\hline
{\bf 1y} &    42.8\% &    34.3\% &    31.0\% &    28.8\% &    27.7\% \\
{\bf 2y} &    28.7\% &    25.6\% &    24.1\% &    23.1\% &    22.4\% \\
{\bf 3y} &    23.5\% &    21.1\% &    20.4\% &    20.0\% &    19.7\% \\
{\bf 4y} &    19.9\% &    18.5\% &    18.2\% &    18.1\% &    18.0\% \\
{\bf 5y} &    17.6\% &    16.8\% &    16.9\% &    16.9\% &    17.0\% \\
{\bf 7y} &    15.4\% &    15.3\% &    15.3\% &    15.3\% &    15.3\% \\
{\bf 10y} &   14.2\% &    14.2\% &    14.2\% &    14.3\% &    14.4\% \\\hline
\end{tabular}}\\\medskip
{\small
\begin{tabular}{|c|ccccc|}\hline
{\bf $\boldsymbol t\downarrow$ / $\boldsymbol b\rightarrow$}   & {\bf 6y} & {\bf 7y} & {\bf 8y} & {\bf 9y} & {\bf 10y} \\\hline
{\bf 1y} &    26.9\% &    26.5\% &    26.3\% &    26.2\% &    26.2\% \\
{\bf 2y} &    22.3\% &    22.2\% &    22.3\% &    22.4\% &    22.4\% \\
{\bf 3y} &    19.7\% &    19.7\% &    19.8\% &    19.9\% &    20.1\% \\
{\bf 4y} &    18.1\% &    18.1\% &    18.2\% &    18.2\% &    18.4\% \\
{\bf 5y} &    16.9\% &    17.0\% &    17.0\% &    17.0\% &    17.1\% \\
{\bf 7y} &    15.3\% &    15.3\% &    15.4\% &    15.5\% &    15.6\% \\
{\bf 10y} &   14.5\% &    14.6\% &    14.7\% &    14.8\% &    15.0\% \\\hline
\end{tabular}}
\end{center}
\end{table}

\begin{table}
\caption{Mid and High risk credit spread parameters}\label{tab:yparam}
\begin{center}
\begin{tabular}{|c|cccc|}\hline
           & $\boldsymbol y_0$ & $\boldsymbol\kappa$ & $\boldsymbol\mu$ & $\boldsymbol\nu$ \\\hline
{\bf Mid}  & 0.01 & 0.80 & 0.02 & 0.20 \\
{\bf High} & 0.03 & 0.50 & 0.05 & 0.50 \\\hline
\end{tabular}
\end{center}
\end{table}

\begin{table}
\caption{Mid risk initial CDS term structure}\label{tab:cdstermmid}
\begin{center}
\begin{tabular}{|c|cccccccccc|}\hline
{$\boldsymbol T$} &    {\bf 1y} &    {\bf 2y} & {\bf 3y} & {\bf 4y} & {\bf 5y} & {\bf 6y} & {\bf 7y} & {\bf 8y} & {\bf 9y} &  {\bf 10y} \\\hline
{\bf CDS Spread} &     92 &        104 &        112 &        117 &        120 &        122 &        124 &        125 &        126 &        127 \\\hline
\end{tabular}
\end{center}
\end{table}

\begin{table}
\caption{Mid risk CDS implied volatility associated to the parameters in Table~\ref{tab:yparam}. Each column contains volatilities of CDS options of a given maturity $T$ for different expiries $t$.}\label{tab:cdsvolsmid}
\begin{center}
{\small
\begin{tabular}{|c|ccccccccc|}\hline
{\bf $\boldsymbol t\downarrow$ / $\boldsymbol T\rightarrow$} &   {\bf 2y} &   {\bf 3y} &   {\bf 4y} &   {\bf 5y} &   {\bf 6y} &   {\bf 7y} &   {\bf 8y} &   {\bf 9y} &  {\bf 10y} \\\hline
  {\bf 1y} &       52\% &       36\% &       27\% &       21\% &       17\% &       15\% &       13\% &       12\% &       11\% \\
  {\bf 2y} &            &       39\% &       28\% &       21\% &       17\% &       14\% &       12\% &       11\% &       10\% \\
  {\bf 3y} &            &            &       33\% &       24\% &       18\% &       15\% &       12\% &       11\% &        9\% \\
  {\bf 4y} &            &            &            &       29\% &       21\% &       16\% &       13\% &       11\% &        9\% \\
  {\bf 5y} &            &            &            &            &       26\% &       19\% &       15\% &       12\% &       10\% \\
  {\bf 6y} &            &            &            &            &            &       24\% &       17\% &       13\% &       11\% \\
  {\bf 7y} &            &            &            &            &            &            &       23\% &       16\% &       12\% \\
  {\bf 8y} &            &            &            &            &            &            &            &       21\% &       15\% \\
  {\bf 9y} &            &            &            &            &            &            &            &            &       19\% \\\hline
\end{tabular}}
\end{center}
\end{table}

\begin{table}
\caption{High risk initial CDS term structure}\label{tab:cdstermhigh}
\begin{center}
\begin{tabular}{|c|cccccccccc|}\hline
{$\boldsymbol T$} &  {\bf 1y} & {\bf 2y} & {\bf 3y} & {\bf 4y} & {\bf 5y} & {\bf 6y} & {\bf 7y} & {\bf 8y} & {\bf 9y} &  {\bf 10y} \\\hline
{\bf CDS Spread}  &   234     &      244 &      248 &      250 &      252 &      252 &      254 &      253 &      254 &        254 \\\hline
\end{tabular}
\end{center}
\end{table}

\begin{table}
\caption{High risk CDS implied volatility associated to the parameters in Table~\ref{tab:yparam}. Each column contains volatilities of CDS options of a given maturity $T$ for different expiries $t$.}\label{tab:cdsvolshigh}
\begin{center}
{\small
\begin{tabular}{|c|ccccccccc|}\hline
{\bf $\boldsymbol t\downarrow$ / $\boldsymbol T\rightarrow$} &   {\bf 2y} &   {\bf 3y} &   {\bf 4y} &   {\bf 5y} &   {\bf 6y} &   {\bf 7y} &   {\bf 8y} &   {\bf 9y} &  {\bf 10y} \\\hline
  {\bf 1y} &       96\% &       69\% &       53\% &       43\% &       36\% &       31\% &       28\% &       26\% &       24\% \\
  {\bf 2y} &            &       71\% &       52\% &       40\% &       32\% &       27\% &       24\% &       21\% &       19\% \\
  {\bf 3y} &            &            &       59\% &       43\% &       33\% &       26\% &       22\% &       20\% &       18\% \\
  {\bf 4y} &            &            &            &       51\% &       37\% &       28\% &       23\% &       20\% &       17\% \\
  {\bf 5y} &            &            &            &            &       45\% &       33\% &       26\% &       21\% &       18\% \\
  {\bf 6y} &            &            &            &            &            &       40\% &       30\% &       24\% &       19\% \\
  {\bf 7y} &            &            &            &            &            &            &       40\% &       29\% &       22\% \\
  {\bf 8y} &            &            &            &            &            &            &            &       36\% &       26\% \\
  {\bf 9y} &            &            &            &            &            &            &            &            &       34\% \\\hline
\end{tabular}}
\end{center}
\end{table}

\section{Pricing Interest-Rate Swaps}\label{app:pricing}

We report in the following tables the detailed numerical results discussed in Section \ref{sec:numerics}. Tables \ref{tab:irs_rec_sym_wwr}, \ref{tab:irs_rec_asym_wwr}, \ref{tab:irs_pay_sym_wwr} and \ref{tab:irs_pay_asym_wwr} shows the prices of non-collateralized bilateral traded IRS with the market setting ''H/M'' for different funding policies and different correlation among market and credit risks. Table \ref{tab:irs_rec_diff} displays the impact of asymmetric rates on bid-ask spreads. Tables \ref{tab:irs_rec_sym_vm_wwr} and \ref{tab:irs_pay_sym_vm_wwr} list the prices of collateralized bilateral traded IRS with the market setting ''H/M'' for different variation margin fractions and different correlation among market and credit risks. Table \ref{tab:irs_rec_sym_decomp} shows the price decompositions for collateralized bilateral and CCP cleared IRS with for different confidence levels for the valuation of initial margin..

\begin{table}
\caption{Prices of a ten-year receiver IRS with unitary notional with high-risk parameter set for the counterparty and mid-risk parameter set for the investor (''H/M``) for different correlations among market and credit risks ($\bar{\rho} = \bar{\rho}^C = \bar{\rho}^I$), and different funding ($\beta^+$) and investing ($\beta^-$) liquidity multipliers. Symmetric funding rates ($\beta^-=\beta^+$). Prices are in basis points.}
\label{tab:irs_rec_sym_wwr}
\begin{center}
\centering
{\small
\begin{tabular}{|c|cccccc|}\hline
 &  \multicolumn{6}{|c|}{\bf Receiver, H/M, $\boldsymbol {\beta^-=\beta^+}$}  \\\cline{2-7}
$\boldsymbol {\bar\rho}\downarrow$ /  $\boldsymbol {\beta^+}\rightarrow$ & {\bf 0\%} & {\bf 20\%} & {\bf 40\%} & {\bf 60\%} & {\bf 80\%} & {\bf 100\%} \\\hline
{\bf -60\%} &    -33.54 & -32.97 & -32.43 & -31.9 & -31.39 & -30.90 \\
{\bf -40\%} &    -21.18 & -20.89 & -20.61 & -20.34 & -20.08 & -19.82 \\
{\bf -20\%} &    -10.18 & -10.16 & -10.14 & -10.12 & -10.10 & -10.07 \\\hline
{\bf 0\%} &       0.28 & 0.05 & -0.17 & -0.38 & -0.58 & -0.77  \\\hline
{\bf 20\%} &     10.36 & 9.89 & 9.45 & 9.02 & 8.61 & 8.22 \\
{\bf 40\%} &     18.01 & 17.35 & 16.73 & 16.13 & 15.55 & 14.99 \\
{\bf 60\%} &     25.18 & 24.36 & 23.57 & 22.81 & 22.09 & 21.39 \\
\hline
\end{tabular}}
\end{center}
\end{table}

\begin{table}
\caption{Prices of a ten-year receiver IRS with unitary notional with high-risk parameter set for the counterparty and mid-risk parameter set for the investor (''H/M``) for different correlations among market and credit risks ($\bar{\rho} = \bar{\rho}^C = \bar{\rho}^I$), and different funding ($\beta^+$) and investing ($\beta^-$) liquidity multipliers. Asymmetric funding rates ($\beta^-=0$). Prices are in basis points.}
\label{tab:irs_rec_asym_wwr}
\begin{center}
\centering
{\small
\begin{tabular}{|c|cccccc|}\hline
 &  \multicolumn{6}{|c|}{\bf Receiver, H/M, $\boldsymbol {\beta^-=0}$}  \\\cline{2-7}
$\boldsymbol {\bar\rho}\downarrow$ /  $\boldsymbol {\beta^+}\rightarrow$ & {\bf 0\%} & {\bf 20\%} & {\bf 40\%} & {\bf 60\%} & {\bf 80\%} & {\bf 100\%} \\\hline
{\bf -60\%} &      -33.54 & -33.86 & -34.17 & -34.48 & -34.78 & -35.08 \\
{\bf -40\%} &      -21.18 & -21.59 & -21.99 & -22.37 & -22.76 & -23.13 \\
{\bf -20\%} &      -10.18 & -10.68 & -11.17 & -11.65 & -12.11 & -12.57 \\\hline
{\bf 0\%} &          0.28 &  -0.33 &  -0.91 &  -1.48 &  -2.04 &  -2.58 \\\hline
{\bf 20\%} &        10.36 &   9.64 &   8.94 &   8.27 &   7.61 &   6.98 \\
{\bf 40\%} &        18.01 &  17.18 &  16.37 &  15.59 &  14.84 &  14.11 \\
{\bf 60\%} &        25.18 &  24.23 &  23.31 &  22.42 &  21.57 &  20.73 \\
\hline
\end{tabular}}
\end{center}
\end{table}

\begin{table}
\caption{Prices of a ten-year payer IRS with unitary notional with high-risk parameter set for the counterparty and mid-risk parameter set for the investor (''H/M``) for different correlations among market and credit risks ($\bar{\rho} = \bar{\rho}^C = \bar{\rho}^I$), and different funding ($\beta^+$) and investing ($\beta^-$) liquidity multipliers. Symmetric funding rates ($\beta^-=\beta^+$). Prices are in basis points.}
\label{tab:irs_pay_sym_wwr}
\begin{center}
\centering
{\small
\begin{tabular}{|c|cccccc|}\hline
 &  \multicolumn{6}{|c|}{\bf Payer, H/M, $\boldsymbol {\beta^-=\beta^+}$}  \\\cline{2-7}
$\boldsymbol {\bar\rho}\downarrow$ /  $\boldsymbol {\beta^+}\rightarrow$ & {\bf 0\%} & {\bf 20\%} & {\bf 40\%} & {\bf 60\%} & {\bf 80\%} & {\bf 100\%} \\\hline
{\bf -60\%} &     -20.72 & -20.70 & -20.67 & -20.63 & -20.58 & -20.52  \\
{\bf -40\%} &     -30.69 & -30.43 & -30.18 & -29.92 & -29.66 & -29.40  \\
{\bf -20\%} &     -40.55 & -40.06 & -39.58 & -39.10 & -38.64 & -38.18  \\\hline
{\bf 0\%} &       -51.08 & -50.33 & -49.6 & -48.89 & -48.19 & -47.51  \\\hline
{\bf 20\%} &      -61.49 & -57.33 & -53.79 & -50.72 & -48.02 & -45.61 \\
{\bf 40\%} &      -74.78 & -73.46 & -72.17 & -70.93 & -69.72 & -68.54 \\
{\bf 60\%} &      -87.83 & -86.17 & -84.57 & -83.02 & -81.52 & -80.06 \\
\hline
\end{tabular}}
\end{center}
\end{table}

\begin{table}
\caption{Prices of a ten-year payer IRS with unitary notional with high-risk parameter set for the counterparty and mid-risk parameter set for the investor (''H/M``) for different correlations among market and credit risks ($\bar{\rho} = \bar{\rho}^C = \bar{\rho}^I$), and different funding ($\beta^+$) and investing ($\beta^-$) liquidity multipliers. Asymmetric funding rates ($\beta^-=0$). Prices are in basis points.}
\label{tab:irs_pay_asym_wwr}
\begin{center}
\centering
{\small
\begin{tabular}{|c|cccccc|}\hline
 &  \multicolumn{6}{|c|}{\bf Payer, H/M, $\boldsymbol {\beta^-=0}$}  \\\cline{2-7}
$\boldsymbol {\bar\rho}\downarrow$ /  $\boldsymbol {\beta^+}\rightarrow$ & {\bf 0\%} & {\bf 20\%} & {\bf 40\%} & {\bf 60\%} & {\bf 80\%} & {\bf 100\%} \\\hline
{\bf -60\%} &     -20.72 & -21.18 & -21.61 & -22.04 & -22.44 & -22.84  \\
{\bf -40\%} &     -30.69 & -31.05 & -31.41 & -31.75 & -32.08 & -32.40  \\
{\bf -20\%} &     -40.55 & -40.84 & -41.12 & -41.39 & -41.66 & -41.92  \\\hline
{\bf 0\%} &       -51.08 & -51.30 & -51.51 & -51.73 & -51.93 & -52.13  \\\hline
{\bf 20\%} &      -61.49 & -61.66 & -61.82 & -61.98 & -62.14 & -62.29 \\
{\bf 40\%} &      -74.78 & -74.90 & -75.01 & -75.13 & -75.24 & -75.35 \\
{\bf 60\%} &      -87.83 & -87.91 & -87.99 & -88.08 & -88.16 & -88.24 \\
\hline
\end{tabular}}
\end{center}
\end{table}

\begin{table}
\caption{Bid-ask spread for a ten-year receiver IRS with unitary notional with high-risk parameter set for the counterparty and mid-risk parameter set for the investor (''H/M``) for different correlations among market and credit risks ($\bar{\rho} = \bar{\rho}^C = \bar{\rho}^I$), and different funding ($\beta^+$) liquidity multipliers. Asymmetric funding rates ($\beta^+\geq\beta^-=0$). Prices are in basis points.}
\label{tab:irs_rec_diff}
\begin{center}
\centering
{\small
\begin{tabular}{|c|cccccc|}\hline
 &  \multicolumn{6}{|c|}{\bf Receiver, H/M, $\boldsymbol {\beta^-=0}$}  \\\cline{2-7}
$\boldsymbol {\bar\rho}\downarrow$ /  $\boldsymbol {\beta^+}\rightarrow$ & {\bf 0\%} & {\bf 20\%} & {\bf 40\%} & {\bf 60\%} & {\bf 80\%} & {\bf 100\%} \\\hline
{\bf -60\%} &     0 & 1.26 & 2.49 & 3.69 & 4.87 & 6.02  \\
{\bf -40\%} &     0 & 1.11 & 2.20 & 3.26 & 4.30 & 5.32  \\
{\bf -20\%} &     0 & 1.01 & 2.00 & 2.97 & 3.91 & 4.84 \\\hline
{\bf 0\%} &       0 & 0.95 & 1.88 & 2.79 & 3.68 & 4.55 \\\hline
{\bf 20\%} &      0 & 0.96 & 1.90 & 2.81 & 3.70 & 4.58 \\
{\bf 40\%} &      0 & 1.01 & 1.99 & 2.95 & 3.88 & 4.78 \\
{\bf 60\%} &      0 & 1.08 & 2.13 & 3.14 & 4.13 & 5.09 \\
\hline
\end{tabular}}
\end{center}
\end{table}

\begin{table}
\caption{Prices of a ten-year receiver IRS with unitary notional with high-risk parameter set for the counterparty and mid-risk parameter set for the investor (''H/M``) for different correlations among market and credit risks ($\bar{\rho} = \bar{\rho}^C = \bar{\rho}^I$), and different fractions of variation margin without initial margin. No margin period of risk. Symmetric funding rates ($\beta^-=\beta^+=1$). Prices are in basis points.}
\label{tab:irs_rec_sym_vm_wwr}
\begin{center}
\centering
{\small
\begin{tabular}{|c|cccccc|}\hline
 &  \multicolumn{6}{|c|}{\bf Receiver, H/M, $\boldsymbol {\beta^-=\beta^+=1}$}  \\\cline{2-7}
$\boldsymbol {\bar\rho}\downarrow$ /  $\boldsymbol {\alpha}\rightarrow$ & {\bf 0\%} & {\bf 20\%} & {\bf 40\%} & {\bf 60\%} & {\bf 80\%} & {\bf 100\%} \\\hline
{\bf -60\%} &     -30.90 & -24.72 & -18.54 & -12.36 & -6.18 & 0.00 \\
{\bf -40\%} &     -19.82 & -15.86 & -11.89 &  -7.93 &  -3.96 & 0.00 \\
{\bf -20\%} &     -10.07 & -8.06 & -6.04 & -4.03 & -2.01 & 0.00 \\\hline
{\bf 0\%} &        -0.77 & -0.62 & -0.46 & -0.31 & -0.15 & 0.00 \\\hline
{\bf 20\%} &        8.22 & 6.58 & 4.93 & 3.29 & 1.64 & 0.00 \\
{\bf 40\%} &       14.99 & 12.00 & 9.00 & 6.00 & 3.00 & 0,00 \\
{\bf 60\%} &       21.39 & 17.11 & 12.83 & 8.56 & 4.28 & 0.00 \\
\hline
\end{tabular}}
\end{center}
\end{table}

\begin{table}
\caption{Prices of a ten-year payer IRS with unitary notional with high-risk parameter set for the counterparty and mid-risk parameter set for the investor (''H/M``) for different correlations among market and credit risks ($\bar{\rho} = \bar{\rho}^C = \bar{\rho}^I$), and different fractions of variation margin without initial margin. No margin period of risk. Symmetric funding rates ($\beta^-=\beta^+=1$). Prices are in basis points.}
\label{tab:irs_pay_sym_vm_wwr}
\begin{center}
\centering
{\small
\begin{tabular}{|c|cccccc|}\hline
 &  \multicolumn{6}{|c|}{\bf Payer, H/M, $\boldsymbol {\beta^-=\beta^+=1}$}  \\\cline{2-7}
$\boldsymbol {\bar\rho}\downarrow$ /  $\boldsymbol {\alpha}\rightarrow$ & {\bf 0\%} & {\bf 20\%} & {\bf 40\%} & {\bf 60\%} & {\bf 80\%} & {\bf 100\%} \\\hline
{\bf -60\%} &     -20.52 & -16.41 & -12.31 & -8.21 & -4.10 & 0.00 \\
{\bf -40\%} &     -29.40 & -23.52 & -17.64 & -11.76 & -5.88 & 0.00 \\
{\bf -20\%} &     -38.18 & -30.54 & -22.91 & -15.27 & -7.64 & 0.00 \\\hline
{\bf 0\%} &       -47.51 & -38.01 & -28.51 & -19.00 & -9.50 & 0.00 \\\hline
{\bf 20\%} &      -56.69 & -45.35 & -34.01 & -22.68 & -11.34 & 0.00 \\
{\bf 40\%} &      -68.54 & -54.84 & -41.13 & -27.42 & -13.71 & 0.00 \\
{\bf 60\%} &      -80.06 & -64.05 & -48.03 & -32.02 & -16.01 & 0.00 \\
\hline
\end{tabular}}
\end{center}
\end{table}

\begin{table}
\caption{Prices of a ten-year receiver IRS traded with a CCP (or bilaterally) with a mid-risk parameter set for the clearing member (investor) and a high-risk parameter set for the client (counterparty) for initial margin posted at various confidence levels $q$. Prices are calculated from the point of view of the client (counterparty). Symmetric funding policy with $\beta^+=\beta^-=1$. Correlation $\bar\rho$ is zero. Prices in basis points with a notional of one Euro.}
\label{tab:irs_rec_sym_decomp}
\begin{center}
\centering
{\small
\begin{tabular}{|c|cccc|cccc|}\hline
 &  \multicolumn{4}{|c|}{\bf Receiver, CCP, $\boldsymbol {\beta^-=\beta^+=1}$} &  \multicolumn{4}{|c|}{\bf Receiver, Bilateral, $\boldsymbol {\beta^-=\beta^+=1}$}  \\\cline{2-9}
$\boldsymbol {q}$ & {\bf CVA} & {\bf DVA} & {\bf MVA} & {\bf FVA} & {\bf CVA} & {\bf DVA} & {\bf MVA} & {\bf FVA} \\\hline
{\bf 50.00\%} & -0.1259 & 3.0798 & 0.0000 & -0.1574 & -2.1317 & 4.3477 & 0.0000 & -0.0842 \\
{\bf 68.00\%} & -0.0660 & 1.6046 & -2.9332 & 0.1251 & -1.1176 & 2.2613 & -4.1389 & 0.2491 \\
{\bf 90.00\%} & -0.0151 & 0.3566 & -8.0373 & 0.5492 & -0.2578 & 0.4997 & -11.3410 & 0.7924 \\
{\bf 95.00\%} & -0.0067 & 0.1540 & -10.3158 & 0.7205 & -0.1149 & 0.2151 & -14.5561 & 1.0250 \\
{\bf 99.00\%} & -0.0011 & 0.0249 & -14.5898 & 1.0290 & -0.0204 & 0.0346 & -20.5869 & 1.4544 \\
{\bf 99.50\%} & -0.0006 & 0.0127 & -16.1545 & 1.1402 & -0.0107 & 0.0176 & -22.7947 & 1.6107 \\
{\bf 99.70\%} & -0.0004 & 0.0083 & -17.2329 & 1.2165 & -0.0070 & 0.0114 & -24.3164 & 1.7184 \\
{\bf 99.90\%} & -0.0002 & 0.0042 & -19.3806 & 1.3684 & -0.0035 & 0.0056 & -27.3469 & 1.9326 \\
\hline
\end{tabular}}
\end{center}
\end{table}

\newpage

\bibliographystyle{plainnat}
\bibliography{ccp_im}

\end{document}